\documentclass[twoside]{article}
\usepackage{qic}

\usepackage{amsmath}
\usepackage{subfigure}

\newcommand\myell{\ell}

\newcommand\beq{\begin{equation}}
\newcommand\eeq{\end{equation}}
\newcommand\bea{\begin{eqnarray}}
\newcommand\eea{\end{eqnarray}}

\begin{document}

\setlength{\textheight}{8.0truein}

\runninghead{A flow-map model for analyzing pseudothresholds $\ldots$}{ K.~M. Svore, A.~W. Cross, I.~L. Chuang, and A.~V. Aho}

\normalsize\textlineskip
\thispagestyle{empty}
\setcounter{page}{1}

\copyrightheading{0}{0}{2005}{000--000}

\vspace*{0.88truein}

\alphfootnote

\fpage{1}

\centerline{\bf A flow-map model for analyzing}
\centerline{\bf pseudothresholds in fault-tolerant quantum computing}
\baselineskip=10pt
\vspace*{0.37truein}
\centerline{\footnotesize
Krysta M. Svore}
\vspace*{0.015truein}
\centerline{\footnotesize\it Columbia University, Dept. of Computer Science, 
1214 Amsterdam Ave. MC:0401} 
\baselineskip=10pt
\centerline{\footnotesize\it New York, NY 10027}
\vspace*{10pt}
\centerline{\footnotesize
Andrew W. Cross}
\vspace*{0.015truein}
\centerline{\footnotesize\it Massachusetts Institute of Technology, Dept. of 
Electrical Engineering, 77 Massachusetts Ave.}
\baselineskip=10pt
\centerline{\footnotesize\it Cambridge, MA 02139}
\vspace*{10pt}
\centerline{\footnotesize
Isaac L. Chuang}
\vspace*{0.015truein}
\centerline{\footnotesize\it Massachusetts Institute of Technology, Dept. of 
Electrical Engineering, 77 Massachusetts Ave.}
\baselineskip=10pt
\centerline{\footnotesize\it Cambridge, MA 02139}
\vspace*{10pt}
\centerline{\footnotesize
Alfred V. Aho}
\vspace*{0.015truein}
\centerline{\footnotesize\it Columbia University, Dept. of Computer Science, 
1214 Amsterdam Ave. MC:0401} 
\baselineskip=10pt
\centerline{\footnotesize\it New York, NY 10027}
\vspace*{0.225truein}
\publisher{(received date)}{(revised date)}

\vspace*{0.21truein}

\abstracts{
An arbitrarily reliable quantum computer can be efficiently constructed from noisy components using
a recursive simulation procedure, provided that those components fail with probability less than the 
fault-tolerance threshold. Recent estimates of the threshold are near some experimentally achieved 
gate fidelities. However, the landscape of threshold estimates includes pseudothresholds, threshold
estimates based on a subset of components and a low level of the recursion. 
In this paper, we observe that pseudothresholds are a generic phenomenon in fault-tolerant computation. 
We define pseudothresholds and present classical and quantum fault-tolerant circuits exhibiting 
pseudothresholds that differ by a factor of $4$ from fault-tolerance thresholds for typical relationships between component failure rates. 
We develop tools for visualizing how reliability is influenced by recursive simulation in order to
determine the asymptotic threshold. Finally, we conjecture that refinements of these methods may 
establish upper bounds on the fault-tolerance threshold for particular codes and noise models.
}{}{}

\vspace*{10pt}

\keywords{Fault-Tolerance}
\vspace*{3pt}
\communicate{to be filled in by the Editorial}

\section{Introduction}\label{sec:intro}
\noindent

A quantum computer can potentially solve certain problems more efficiently 
than a classical computer \cite{shor94,grover97,hallgren02}.
However, quantum computers are likely to be engineered from inherently noisy 
components, so any scalable quantum computer system will require quantum error 
correction and fault-tolerant methods of computation.
As candidate quantum device technologies mature, we need to determine
component failure probabilities necessary to achieve scalability. 
The fault-tolerance threshold for gate and memory components is particularly 
interesting because arbitrarily reliable computations are possible if the 
circuit components have failure rates below the threshold. Given detailed knowledge 
of the fault-tolerance threshold and its associated trade-offs, proposals 
for fault-tolerant quantum computation can be critically evaluated.

The concept of a fault-tolerance threshold has its origins in the classical theory of
computation. In the 1950's, von Neumann showed that it is possible to achieve a reliable 
classical computation with faulty components provided that the failure 
probability of each component is below some constant threshold probability 
that is independent of the circuit size and the desired noise rate 
\cite{vonneumann56}.
Similarly, concatenated coding and recursive error correction can be used to achieve 
reliable quantum computation. 
Concatenation is the process of encoding physical bits of one code as logical bits of another code.
It is now well-known that using a 
single-error-correcting concatenated coding scheme with $L$ levels of recursion, 
the maximum failure probability $\gamma_{circuit}$ of a fault-tolerant circuit 
can be estimated as a function of the maximum failure probability $\gamma$
of a basic component using the {\it fault-tolerance threshold inequality}
\beq
\frac{\gamma_{circuit}(\gamma)}{\gamma_{th}} \leq \left(\frac{\gamma}{\gamma_{th}}\right)^{2^L},
\label{eqn:thresh}
\eeq
where $\gamma_{th}$ is the asymptotic threshold \cite{preskill97}.
When Eq~(\ref{eqn:thresh}) holds with equality, we call it the
{\it fault-tolerance threshold equation}. The final circuit failure probability 
$\gamma_{circuit}$ decreases as a doubly exponential function of $L$ if $\gamma<\gamma_{th}$.

\begin{figure}[htbp]
\vspace*{13pt} \centerline{\psfig{file=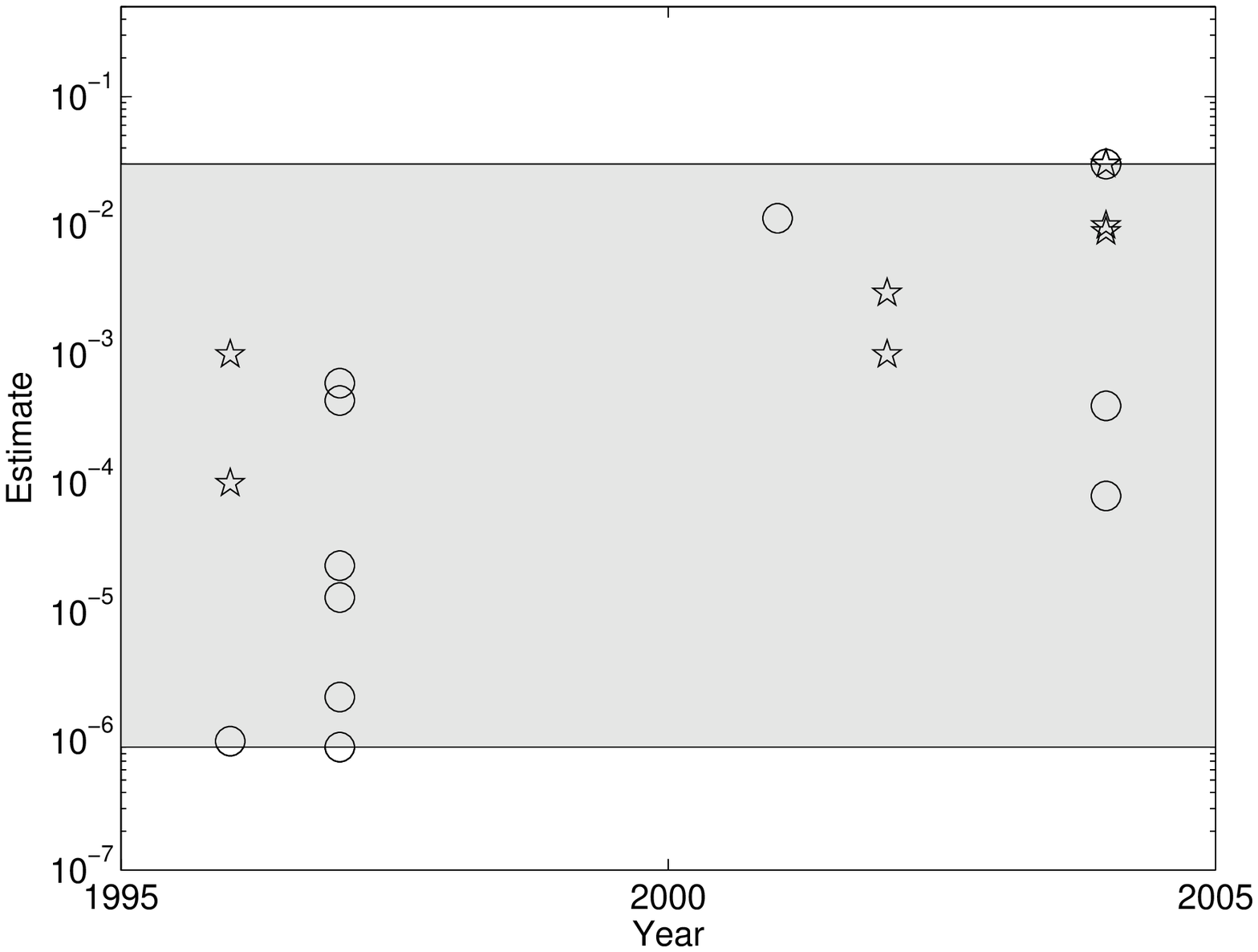,width=8.2cm}}
\vspace*{13pt} \fcaption{Plots of quantum threshold estimates for
stochastic noise models between the years 1996 and 2004
\cite{preskill97,svore04local,aharonov99, dennis01, gottesmanthesis, knill97, knill04,ohno04,reichardt04,steane02overhead,zalka96}. 
Stars denote numerical estimates and circles denote all others.
Most of these
estimates apply to the $[[7,1,3]]$ code, though the pair of thresholds at and
above $10^{-2}$ apply to surface code memories and post-selected quantum 
computation. The networks vary over unitary networks, 
nearest-neighbor networks, and various optimized networks. Some estimates apply only when 
there are no memory errors, and others apply only for the Clifford-group gates.
The dark swath designates the large interval that contains all of these 
estimates.} 
\label{fig:thresh}
\end{figure}

One branch of fault-tolerant quantum computing research has focused on 
estimating the fault-tolerance threshold in the fault-tolerance threshold 
inequality. 
Figure~\ref{fig:thresh} shows the range of quantum threshold estimates
reported in 
\cite{preskill97,svore04local,aharonov99, dennis01, gottesmanthesis, knill97, knill04,ohno04,reichardt04,steane02overhead,zalka96}.
The estimates, which vary between $10^{-6}$ and $10^{-2}$,
include numerical and analytical results for varying networks for the $[[7,1,3]]$ code, such as optimized networks, unitary networks, and nearest-neighbor networks. They also include
results for surface code memories and post-selected quantum computing
models that yield thresholds above $10^{-2}$. Numerical estimates tend
to be more optimistic than their analytical counterparts.

\begin{figure}[htbp]
\vspace*{13pt} \centerline{ \subfigure{
\psfig{file=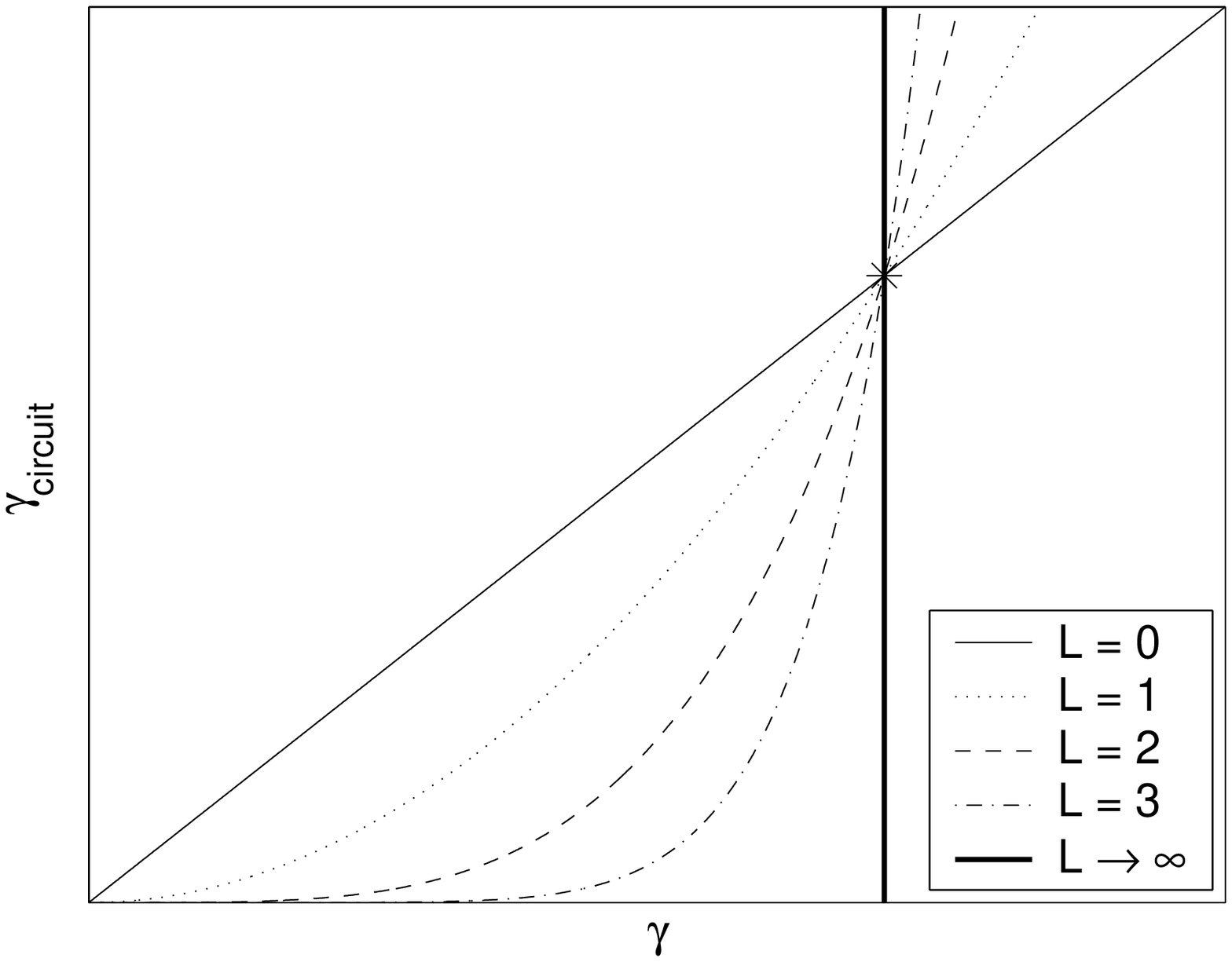,width=5cm} \label{fig:trp:a} } \hspace{1cm}
\subfigure{ \psfig{file=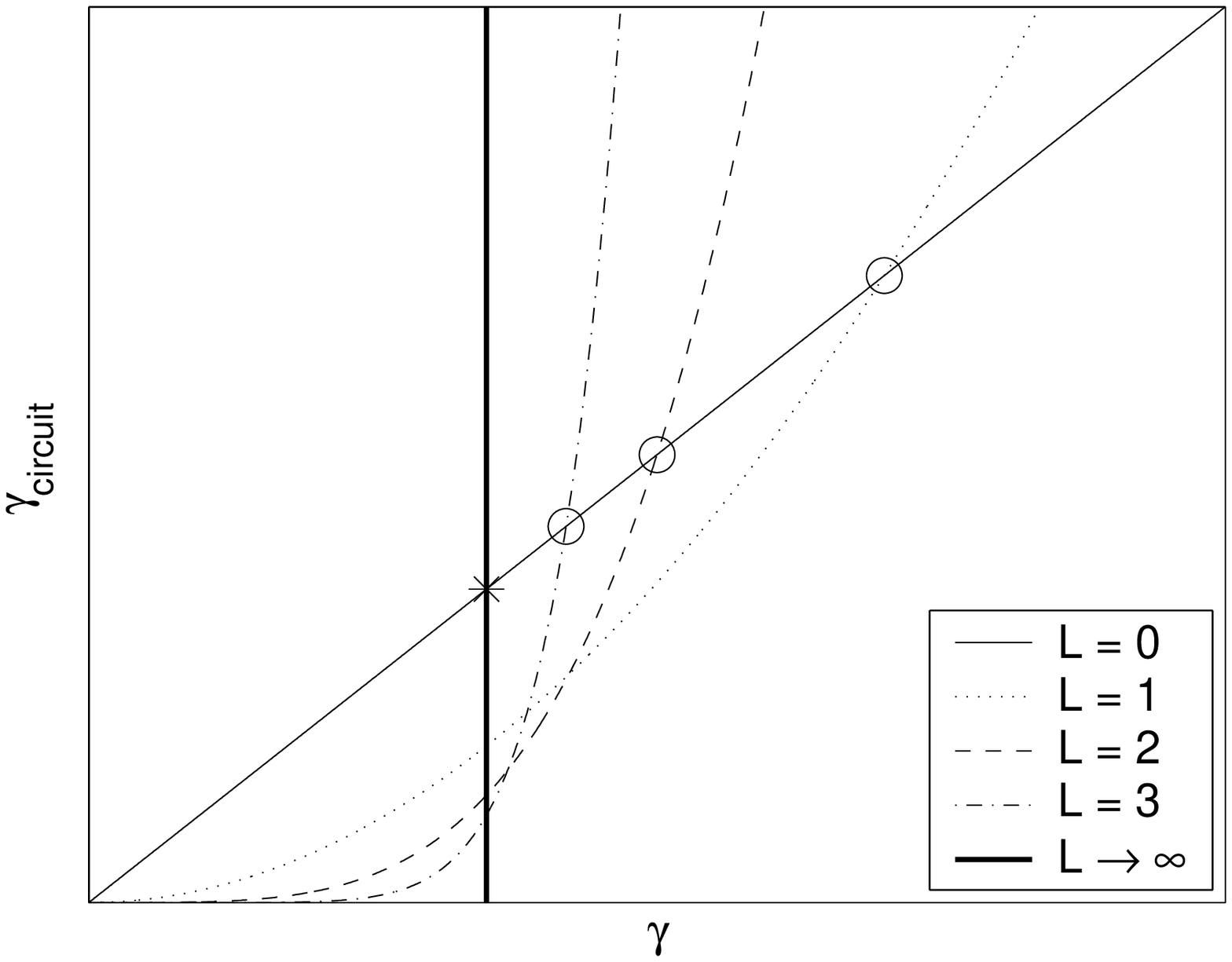,width=5cm} \label{fig:trpb:b} }}
\fcaption{(a) An ideal threshold reliability information plot (TRIP)
follows Eq~(\protect{\ref{eqn:thresh}}). The crossing point between
the $L=0$ line and the $L=1$ curve, marked by the asterisk on the thick
vertical line, is the fault-tolerance threshold. All of the curves
cross at the same point. (b) A real TRIP does not follow
Eq~(\protect{\ref{eqn:thresh}}). The crossing points between the
$L=0$ line and the other curves are all different. These points,
marked by circles, correspond to a sequence of pseudothresholds that
converges to the real fault-tolerance threshold marked by the
asterisk on the thick vertical line.}
\end{figure}

If there is a single type of component that is replaced using the same rule each time, the fault-tolerance threshold inequality, 
Eq~(\ref{eqn:thresh}), becomes an equality. This leads to a very simple prediction 
shown in Figure~\ref{fig:trp:a}: if $\gamma=\gamma_{th}$, then $\gamma_{circuit}=\gamma_{th}$.
This fact is taken as the basis of several numerical analyses of the fault-tolerance threshold today. 
Specifically, this simplification is attractive for computationally expensive numerical simulations
because it implies that the threshold can be determined by finding the smallest nonzero value of 
$\gamma$ that solves $\gamma_{circuit}(\gamma)=\gamma$.

Realistically, however, there are multiple types of components that are each replaced using
different rules, so the first crossing point does not 
accurately indicate the fault-tolerance threshold. Figure \ref{fig:trpb:b} more accurately 
portrays the effect of recursive simulation. As the recursion level increases, for example, an exponentially 
growing number of wires must be introduced between gates. When these wires are unreliable, as they likely 
will be in quantum circuits, successive recursion levels can cause errors to increase even though 
$\gamma$ is beneath the apparent threshold. Thus, recursive simulation changes the relative proportions 
of each type of component and what appears to be the threshold at one level of recursion may be far from 
the asymptotic threshold. This sequence of crossing points cannot be used to describe the proper conditions
under which a system is scalable. Rather, these crossing points are pseudothresholds \cite{svore04local}.
 
When multiple types of components are replaced using different rules, each component type must be
parameterized by a separate failure probability. Hence, opportunity exists for
engineering trade-offs that still preserve scalability. These trade-offs can be quantified given
the asymptotic threshold, families of pseudothresholds, and their relation to the shape of the set of
subthreshold component parameters.

In this paper, we present practical methods for distinguishing pseudothresholds from 
asymptotic thresholds. In particular, we explore the conditions under which pseudothresholds exist
and clarify their meaning. We embark on this exploration carrying two tools. The first tool is a generalization of Figure~\ref{fig:trpb:b}, 
which we call the {\it threshold reliability information plot} (TRIP). In a TRIP, each curve represents the 
failure probability of a particular component at concatenation level $L$ and crosses the $L=0$ line once. The crossing of a level-$L$ curve and the level-($L=0$) line yields the rightmost edge of an interval on the $\gamma$--axis below which  
reliability is improved by concatenation. The crossing point is a level-$L$ pseudothreshold.

\begin{figure}[htbp]
\vspace*{13pt} \centerline{ \psfig{file=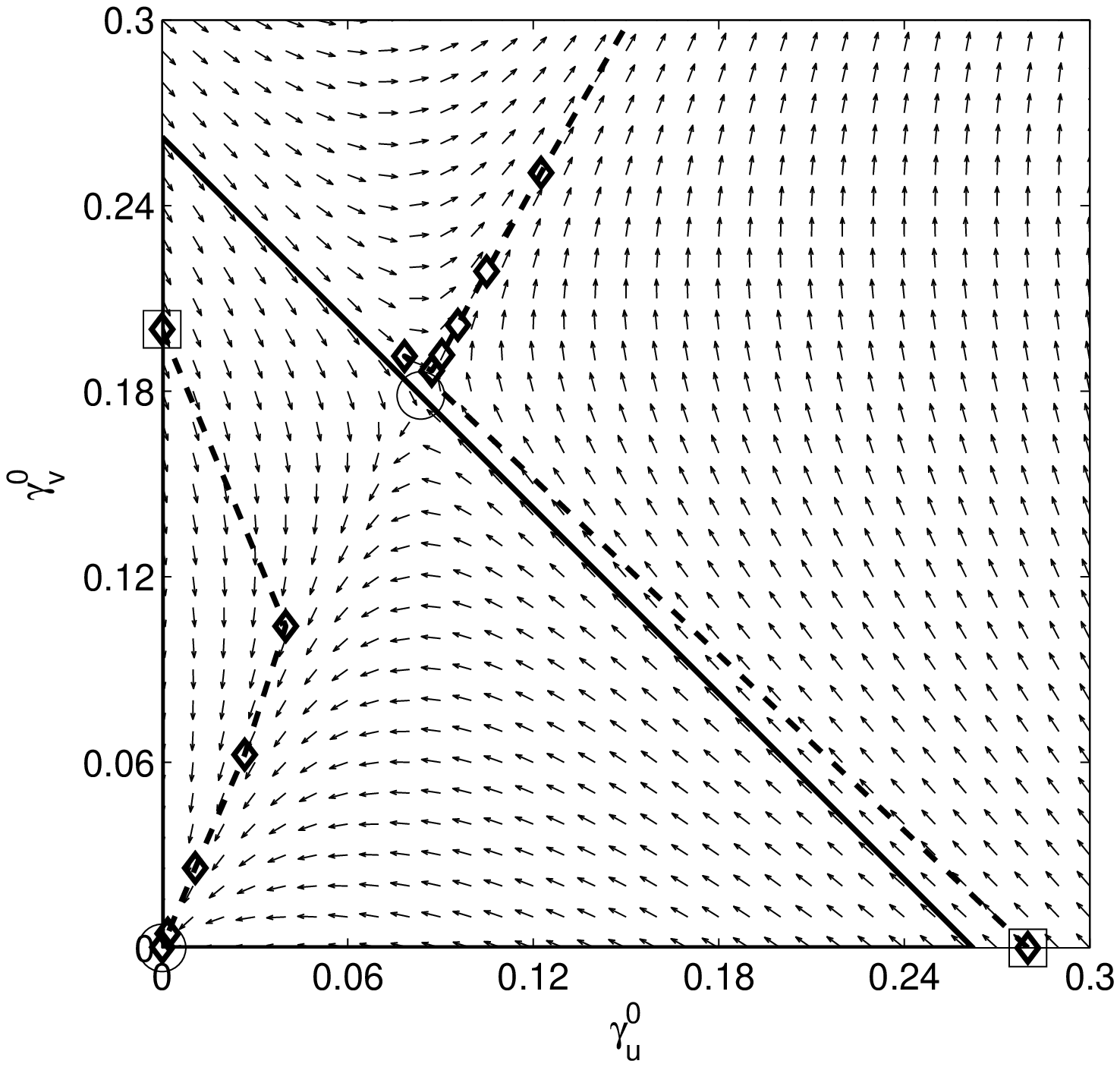,width=8.2cm}}
\vspace*{13pt} \fcaption{ Threshold information flow diagram (TIFD)
corresponding to the following recursive simulation procedure: gates
$u$ and $v$ at level-$(L-1)$ are both replaced by fault-tolerant
gates at level $L$ that can withstand a single level-$(L-1)$ gate
failure. However, $u$ and $v$ compute different functions, so their
fault-tolerant implementations are different. In this example, the
fault-tolerant $u$ contains two $u$ gates and two $v$ gates, and the
fault-tolerant $v$ contains three $u$ gates and three $v$ gates.
Arrows on this TIFD indicate how the recursive simulation procedure
changes the effective failure probabilities of $u$ and $v$. For some
failure probabilities, recursive simulation has no effect. These
fixed points are marked by circles, and one of them determines the
fault-tolerance threshold, indicated by a thick black line. Sample
trajectories begin at the squares and flow along the thick dashed
lines, where the diamonds mark the sequence of points for each level
$L$ of concatenation. } \label{fig:tifd}
\end{figure}

The second tool is a {\it threshold information flow diagram} (TIFD) that shows how
recursive simulation can change the reliability of a particular set of noisy components.
A {\it flow} is a normalized vector field that can be visualized as a collection of arrows.
Each arrow's base is anchored to a point that represents the current failure probability of 
the (composite) components. The direction of each arrow indicates the direction the anchor point moves
at the next level of recursion. In contrast to the TRIP, the TIFD exposes how all of the
component failure probabilities change in a recursive simulation. If the recursive
simulation is self-similar, i.e., if the failure probability of a level-$L$ component can
be expressed in terms of the failure probabilities of level-$(L-1)$ components, particularly 
with respect to the noise model, then the TIFD indicates whether or not recursive simulation 
increases or decreases each component's failure probability, allowing us to compute and
visualize the fault-tolerance threshold.

For example, Figure~\ref{fig:tifd} shows a TIFD for a hypothetical pair of faulty basic gates $u$ and 
$v$. Because $u$ and $v$ compute different functions, 
their fault-tolerant implementations use different numbers of basic $u$ and $v$ gates. In this 
example, the fault-tolerant $u$ contains two $u$ gates and two $v$ gates, connected in some 
fashion so that the fault-tolerant $u$ can withstand one internal gate failure and still produce 
a ``good'' output. If basic gates $u$ and $v$ fail independently with probabilities
$\gamma_u$ and $\gamma_v$, respectively, then the failure probability of the fault-tolerant $u$ 
gate is
\begin{equation}
1 - (1-\gamma_u)^2(1-\gamma_v)^2 - 2\gamma_u(1-\gamma_u)(1-\gamma_v)^2 - 2\gamma_v(1-\gamma_v)(1-\gamma_u)^2.
\end{equation}
Similarly, the fault-tolerant $v$ gate contains three $u$ gates and three $v$ gates 
and can withstand any single failure, giving
\begin{equation}
1 - (1-\gamma_u)^3(1-\gamma_v)^3 - 3\gamma_u(1-\gamma_u)^2(1-\gamma_v)^3 - 3\gamma_v(1-\gamma_v)^2(1-\gamma_u)^3.
\end{equation}
There are many such hypothetical examples. In Section~\ref{sec:classical}, we give a more realistic 
example that demonstrates how a TIFD is calculated based on an actual circuit.

Continuing with this example, the effective failure probabilities of $u$ and $v$ 
after level-$1$ recursive simulation both 
depend on the initial failure probabilities $\gamma_u^0$ and $\gamma_v^0$ of $u$ and $v$,
shown on the horizontal and vertical axes of the TIFD. Consider the following scenario.
The $v$ gate initially fails with probability $0.2$ and the $u$ gate does not fail at all.
A square marks this point on the TIFD. The arrow at this point on the TIFD points down
and to the right, indicating that a level-$1$ recursive simulation will improve $v$ but 
make $u$ worse. The dashed line connects the initial failure probabilities to the 
failure probabilities of the level-$1$ simulated gates (at about $(0.04,0.1)$). The dashed 
path shows that subsequent recursive simulation makes $u$ and $v$ arbitrarily reliable.

The TIFD in Figure~\ref{fig:tifd} also indicates the set of initial failure probabilities that
is below threshold. The boundary of this set is determined by a fixed point of the recursive
simulation procedure. This fixed point is marked with a circle, and the thick dark line
passing through this circle is the invariant set that indicates the fault-tolerance threshold.
For example, an ideal $v$ gate and a $u$ gate that initially fails with probability $0.28$
(marked by a square) is above threshold. This point flows nearly parallel to the invariant
line at first, but ultimately escapes from the boundary of the TIFD after about $7$
levels of recursive simulation.

We organize the paper as follows. In Section \ref{sec:defs}, we first
define the fault-tolerance threshold for concatenated flow maps, then
we define pseudothresholds and describe their importance. We calculate, 
in Section~\ref{sec:classical}, a family of pseudothresholds and distinguish them from the 
fault-tolerance threshold estimate for the classical repetition code.
In Section~\ref{sec:quantum}, we study the $[[7,1,3]]$ CSS code, again comparing 
the fault-tolerance threshold against families of pseudothresholds.
We suggest techniques for finding the threshold in Section~\ref{sec:technique} 
that expand upon our use of the TIFD. We conclude in 
Section~\ref{sec:conclusion} with open questions.

\section{Pseudothresholds}\label{sec:defs}
\noindent

In Section~\ref{sec:intro}, we discussed the phenomenon of pseudothresholds
without introducing many mathematical concepts. In this section, we clarify 
what we mean by defining both the fault-tolerance threshold and the 
pseudothreshold sequences for a set of flow maps.

\subsection{The fault-tolerance threshold}
\noindent

The threshold is related to the number of ``bad'' fault paths 
through a circuit. Assume faults occur at discrete locations and times with probability dependent on 
the location type. The failure probability of the circuit can then be determined
as a function of failure probabilities of types of locations. For a code correcting $t$ 
errors, the probability that one of these bad fault paths occurs is no more than $C\gamma^{t+1}$, 
where $\gamma$ is the largest failure probability of any type of location in the circuit 
and $C$ is the number of ways to choose $t+1$ failed locations out of $N$ total locations.
The fault-tolerance threshold satisfies 
\beq
\label{eq:tbound}
\gamma_{th}\geq \left(\frac{1}{C}\right)^{1/(t+1)},
\eeq
since $\gamma_{th}$ is the fixed point of the map $\gamma_{circuit}(\gamma)=C\gamma^{t+1}$.

However, we must recognize that the fault-tolerant implementations of each 
location type differ, so we must express the failure probability of each 
fault-tolerant location type as a function of the location types it contains. 
In other words,
we construct (approximations to) the {\it flow maps} for the given fault-tolerant implementations and 
noise model \cite{svore04local,gottesmanthesis}. In particular, if each type of location $\myell$ is assigned 
an initial failure probability $\gamma_\myell^0$ and if there are $n$ different types of locations, the approximate 
failure probability of location type $\myell$ after one level of recursive simulation is a function $\Gamma_\myell^1$ 
of all $n$ of the initial failure probabilities.
Therefore, $\Gamma_\myell^1$ is called the {\it flow map for location type $\myell$}.

Considering all types of locations $l$, the functions $\Gamma_\myell^1$ are the coordinates of a {\it flow map} $\Gamma^1$.
The flow map takes the failure probabilities of the $n$ location types to their new values
after one level of recursive simulation. The failure probabilities of the $L$-simulated location 
types are (approximately) related to the initial failure probabilities $\gamma_\myell^0$ by the 
composed flow map, 
\begin{equation}\label{eq:selfsim}
\Gamma^L \approx \underset{\hbox{{\it L} times}}{\underbrace{\Gamma^1\circ\dots\circ\Gamma^1}}.
\end{equation}
Ideally, the replacements can be constructed so that Eq~(\ref{eq:selfsim}) is an
equality. This is the case for the example in Section~\ref{sec:classical} but not in Section~\ref{sec:quantum}.
Let $\Gamma^L_\myell$ denote the coordinate function of $\Gamma^L$ associated with
location $\myell$, and let $\Gamma^0_\myell$ be the initial function that selects the $\myell$
coordinate. In other words, $\Gamma^L_\myell$ is the failure probability
of location $\myell$ after $L$ levels of recursive simulation.
The function $\Gamma^L_\myell$ is a {\it concatenated flow map for location type $\myell$} because we can use
Eq~(\ref{eq:selfsim}) to derive (an approximation to) $\Gamma_\myell^L$. The map $\Gamma^L$ 
is the {\it concatenated flow map}. 

A vector of failure probabilities $\vec{\gamma}$ for the $n$ location types is
{\it below threshold} if all $n$ of the failure probabilities approach zero
as the concatenation level approaches infinity,
\begin{equation}
\lim_{L\rightarrow\infty} \Gamma^L(\vec{\gamma})=\vec{0}.
\end{equation}
Let $T$ be the set of these vectors that are below threshold, and let $C_\epsilon$
be the $n$-dimensional cube of edge length $\epsilon$ with one corner at the origin,
\beq
C_\epsilon\equiv \{\vec{\gamma}\in [0,1]^n\ |\ \gamma_\myell<\epsilon\ \ \forall \myell, \epsilon\in (0,1]\}
\eeq
The cube contains all of the vectors whose worst failure probability is less than
$\epsilon$. The {\it fault-tolerance threshold} or {\it asymptotic threshold} is the 
size of the largest cube contained in $T$, i.e.,
\begin{equation}
\gamma_{th} \equiv \sup \{ \epsilon\geq 0\ |\ C_\epsilon\subseteq T\}.
\end{equation}
If all component failure probabilities are beneath this probability, then
composing the flow maps reduces the failure probability arbitrarily 
close to zero.

\subsection{Definition of pseudothresholds}
\noindent

Before we define pseudothresholds, we introduce the concept of a {\it setting}.
Settings parameterize a set of location failure probabilities by a single
parameter so that we can think of $\Gamma^L_{\myell}$ as a function of
this parameter. A setting is a function from a single failure probability 
parameter to a vector of $n$ failure probabilities, one for each location. 
For example, the {\it diagonal setting} $g(\gamma) = (\gamma,\dots,\gamma)$ 
is used in analyses that assign each location the same initial failure probability.
The {\it Steane setting} $g(\gamma)$ is another setting that sets all location 
failure probabilities to $\gamma$ except for a waiting bit which is assigned $\gamma/10$ 
\cite{steane02overhead} (in this reference, a waiting bit is assigned a
failure probability of $\gamma/100$).

Suppose there are $n$ types of locations. We define a pseudothreshold 
$\gamma_{\myell,g}^L$ for a 
fixed level of recursion $L>0$, a location $\myell$, and a setting $g$ as the
least nonzero solution to
\beq
\Gamma^L_{\myell}(g(\gamma)) = \gamma,
\label{eq:pseudodef}
\eeq
This definition presents the $(L,\myell,g)$-pseudothreshold as a fixed-point 
calculation for a function derived from the flow map $\Gamma$. The left-hand size of 
Eq~(\ref{eq:pseudodef}) can be viewed as one of the curves plotted in 
Figure~\ref{fig:trp:a}, and the right hand size can be viewed as the $L=0$ 
line. The point where these curves intersect is a pseudothreshold. 

For fixed location $\myell$ and setting $g$, the sequence $\gamma^L_{\myell,g}$ 
is not necessarily constant as a function of $L$. In fact, the sequence is
typically not constant, meaning that pseudothresholds are a generic phenomenon.
More specifically, let $\gamma_0$ be any pseudothreshold of the flow map 
$\Gamma$ for a setting $g$, and let $\vec{\gamma_0}$ be the constant vector
$(\gamma_0,\dots,\gamma_0)$. The pseudothreshold $\gamma_0$  is independent
of location type and recursion level only if the setting satisfies
$\Gamma(g(\gamma_0))=\vec{\gamma}_0$ and $\vec{\gamma}_0$ is a fixed 
point of $\Gamma$.

Despite the fact that pseudothresholds are not thresholds, pseudothresholds 
are interesting because only a fixed level of recursive simulation will be 
used in practice. If the goal is to construct a reliable fault-tolerant
location type $\myell$, and all of the location types have the same initial reliability 
(i.e., $g(\gamma)=(\gamma,\dots,\gamma)$), then choosing $\gamma$ to
be less than the $(L,\myell,g)$-pseudothreshold makes the $L$-simulated gate 
location type $\myell$ more reliable than the initial gate. However, some caution must 
be applied to pseudothresholds as well because the $(1,\myell,g)$-pseudothreshold and 
the $(2,\myell,g)$-pseudothreshold can be substantially different.

In the following sections, we present an illustrative example of classical 
pseudothresholds followed by a more detailed example of quantum 
pseudothresholds. We show by means of these examples that pseudothresholds 
are generic to all multiparameter maps.  In addition, we highlight that 
threshold estimates should account for multiple location types and higher levels 
of code concatenation to achieve more realistic threshold results. 

\section{Classical Pseudothresholds for the $[3,1,3]$ Code}\label{sec:classical}
\noindent

In this section, we analyze a classical example to build intuition about the 
differences between pseudothresholds and thresholds.
We study pseudothresholds for classical fault-tolerant
components based on the $[3,1,3]$ repetition code. We use the threshold 
reliability information plot (TRIP) of the $[3,1,3]$ code to identify 
pseudothresholds. We then characterize the flow map for this example 
using a threshold information flow diagram (TIFD).  

\subsection{The $[3,1,3]$ code and its failure probability map}
\noindent

In this example, the classical single-error-correcting $[3,1,3]$ repetition code, also 
called triple modular rendundancy (TMR), is used to encode a single bit in three bits 
by copying it three times. To make a fault-tolerant classical wire using this 
code, three location types $\Omega = \{w,v,f\}$ are required, where
\begin{itemize}
\item $w:\{0,1\}\rightarrow\{0,1\}$ defined by $w(a)=a$ is a {\it wire}.
\item $v:\{0,1\}^3\rightarrow\{0,1\}$ defined by $v(a,b,c)=ab\oplus bc\oplus ca$ is a {\it voter}.
\item $f:\{0,1\}\rightarrow\{0,1\}^3$ defined by $f(a)=(a,a,a)$ is a {\it fanout}.
\end{itemize}
The superscripts above indicate the Cartesian product.
The wire $w$ is analogous to a waiting bit in a quantum fault-tolerance 
analysis, and the voter $v$ and fanout $f$ perform error correction.

A noisy version of each location type is defined as follows. A noisy wire flips the 
output bit with probability $\gamma_w$. A noisy voter incorrectly indicates 
the output bit with probability $\gamma_v$. For simplicity, we choose to model 
the fanout gate to be noiseless. 

To recursively construct a fault-tolerant wire, {\it replacement rules} are used.
A replacement rule is a pair $(b,R(b))$ where $b\in \Omega$ and $R(b)$ is a 
circuit over $\Omega$ that specifies how to replace a level-($L-1$) location at 
level $L$. $R(b)$ is called the {\it replacement} of $b$ and must preserve the 
functionality of the original location $b$. We construct the replacement rules to 
mirror replacement rules for quantum circuits, in which a circuit location is replaced 
by error correction followed by a fault-tolerant implementation of the location. 

The following steps suggest how to ensure proper component connectivity for a code
encoding a single bit. First, replace $b$ directly by 
$D^{\otimes n_o} \circ R(b) \circ E^{\otimes n_i}$ where $D$ and $E$ 
are an ideal decoder and encoder. We must have $D\circ E$ equal to the identity gate on 
a single bit, where the open circle $\circ$ denotes function composition. The numbers
$n_i$ and $n_o$ are the number of input and output bits of $b$, respectively.
After replacing each component in this manner, make a second pass over the circuit 
and replace all pairs $D\circ E$ by bundles of wires. Finally, replace the remaining 
encoders and decoders by respective fault-tolerant implementations of input preparation 
and output readout. The resulting circuit components will be properly connected, and 
the circuit will not contain decoding and re-encoding components since these components 
are not typically fault-tolerant.

\begin{figure}[htbp]
\vspace*{13pt} \centerline{
\psfig{file=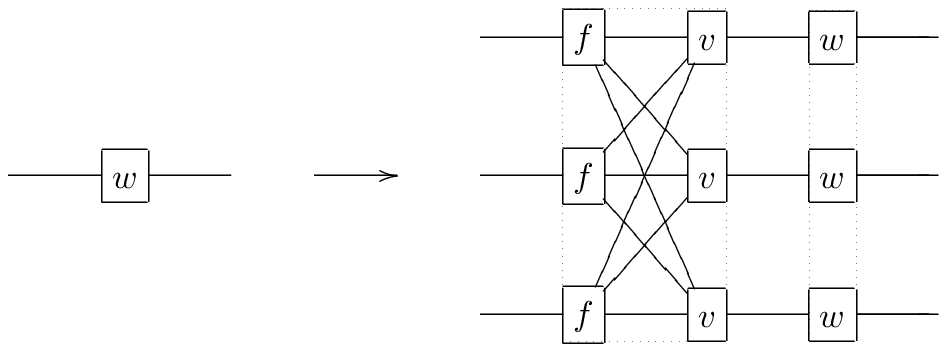,width=8.2cm}}\vspace*{13pt}
\fcaption{Replacement rule for a wire $w$. The fanouts $f$ and
voters $v$ perform error correction. The first dashed box indicates
classical error correction using fanouts $f$ and voters $v$. The
second dashed box indicates the fault-tolerant implementation of the
wire $w$.} \label{fig:wire}
\end{figure}

\begin{figure}[htbp]
\vspace*{13pt}\centerline{ \subfigure{
\psfig{file=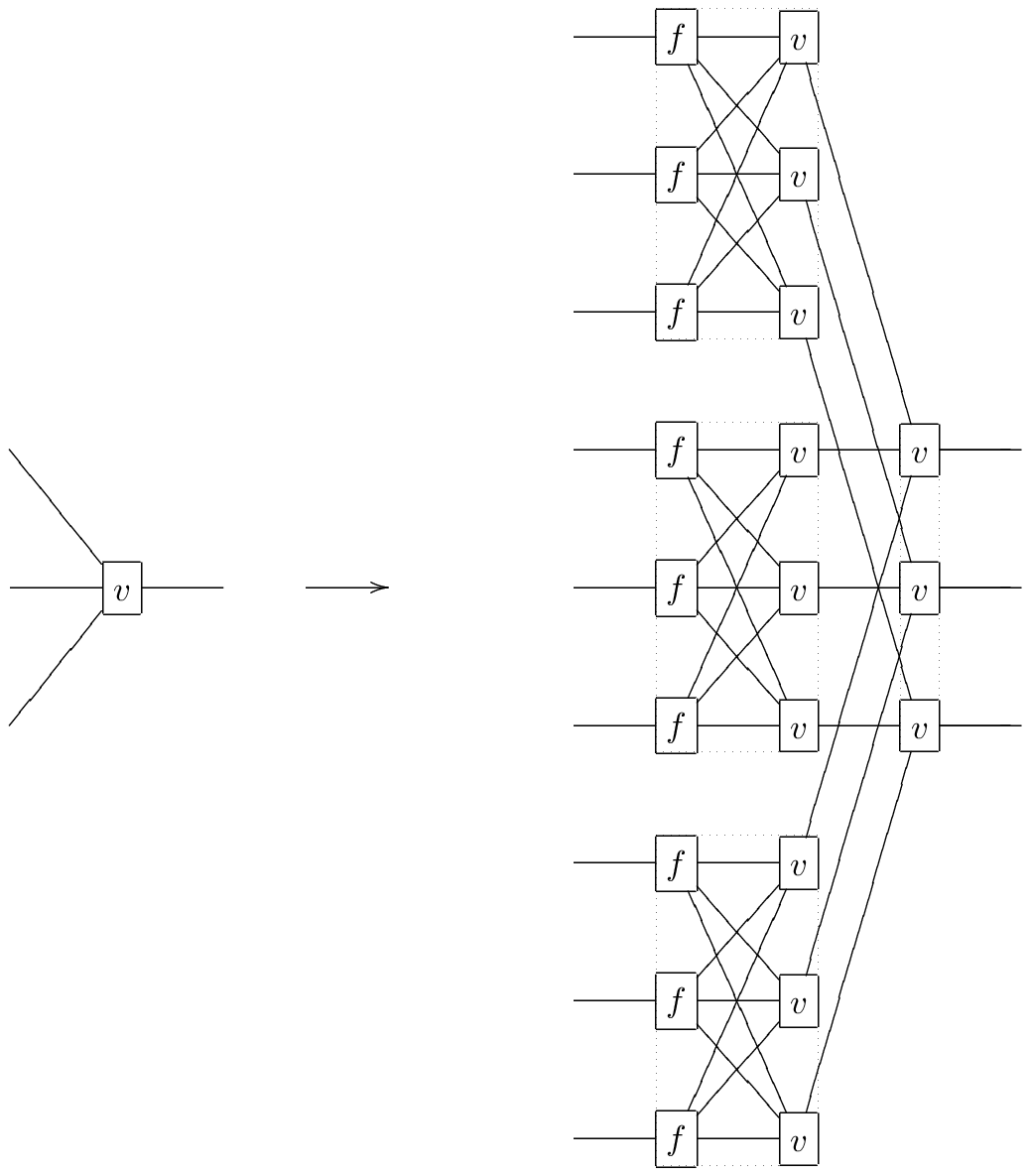,width=4.1cm} \label{fig:voter:a} }
\hspace{1cm} \subfigure{ \psfig{file=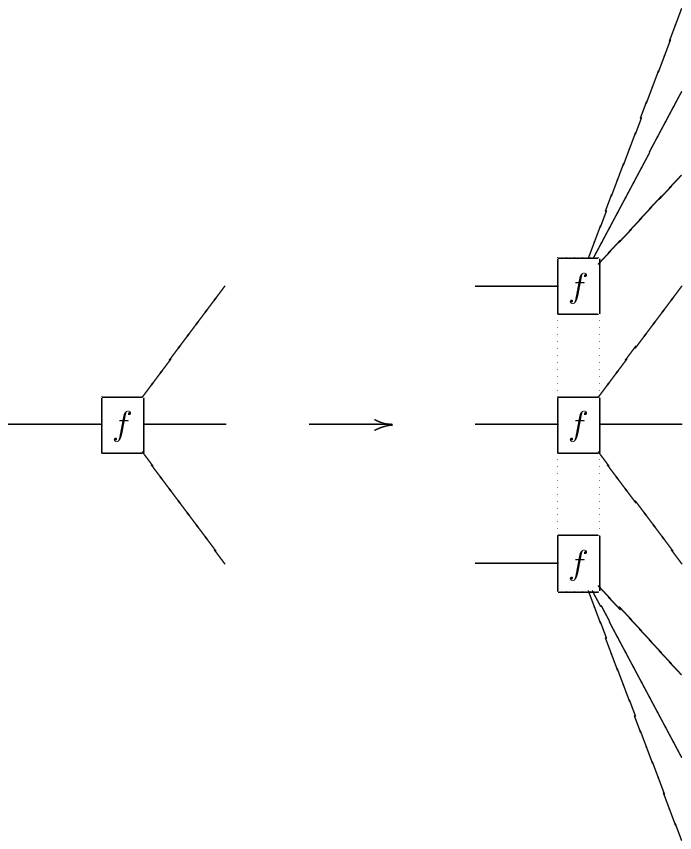,width=4.1cm}
\label{fig:fanout:b} }}\vspace*{13pt} \fcaption{(a) Replacement rule
for a voter $v$. The first dashed box indicates classical error
correction using fanouts $f$ and voters $v$. The second dashed box
indicates the fault-tolerant implementation of the voter $v$. (b)
Replacement rule for a fanout $f$. Because we assume fanouts are
noiseless, the replacement is just three fanout gates $f$.
}
\end{figure}

For this example, a wire $w$ is replaced by error correction 
followed by a transversal implementation of the wire, i.e., a wire is applied 
to each bit of the encoded input, shown in Figure~\ref{fig:wire}. Note the 
first dashed box indicates the classical error correction, which involves 
$w,v,$ and $f$ location types, and the second dashed box indicates the 
fault-tolerant implementation of the original location. Similarly, Figures 
\ref{fig:voter:a} and \ref{fig:fanout:b} show the fault-tolerant replacement 
of $v$ and $f$, respectively.

Using the replacements $R(w)$ and $R(v)$, the failure probabilities $\Gamma_w^1(\vec{\gamma})$ 
and $\Gamma_v^1(\vec{\gamma})$ can be found, where the initial vector of failure probabilities 
is $\vec{\gamma} = (\gamma_w^0,\gamma_v^0)$.  Failure is defined to occur when the component's
output does not decode to the correct value. The wire failure probability is 
easily calculated by counting the number of ways each configuration of errors 
occurs. For example, three voters fail in one way, three voters and one wire 
fail in three ways, two voters fail in three ways, etc. Rewriting the resulting
polynomial in distributed form gives
\begin{align}
\Gamma_w^1(\vec{\gamma}) & = 6\gamma_v\gamma_w + 3\gamma_v^2 + 3\gamma_w^2 - 2\gamma_v^3 - 18\gamma_v^2\gamma_w 
+ 12\gamma_v^3\gamma_w - 18\gamma_v\gamma_w^2 \nonumber \\
& + 36\gamma_v^2\gamma_w^2 - 24\gamma_v^3\gamma_w^2 - 2\gamma_w^3 
+ 12\gamma_v\gamma_w^3 - 24\gamma_v^2\gamma_w^3 + 16\gamma_v^3\gamma_w^3,
\label{eqn:wire}
\end{align}
where the superscript $0$ has been dropped for notational convenience.

Replacing the wire failure probability $\gamma_w^0$ by $\gamma_v^0$ and 
the voter failure probability $\gamma_v^0$ by the probability of error 
correction failure $3(\gamma_v^0)^2(1-\gamma_v^0)+(\gamma_v^0)^3$ gives 
\begin{align}
\Gamma_v^1(\vec{\gamma}) & = 3\gamma_v^2 + 16\gamma_v^3 - 39\gamma_v^4 - 126\gamma_v^5 + 474\gamma_v^6 
- 288\gamma_v^7 - 936\gamma_v^8 \nonumber \\
& + 2080\gamma_v^9 - 1824\gamma_v^{10} + 768\gamma_v^{11} - 128\gamma_v^{12},
\label{eqn:voter}
\end{align}
where the superscript has been dropped again in the last expression.

The flow maps Eq~(\ref{eqn:wire}) and Eq~(\ref{eqn:voter}) are
exact and satisfy Eq~(\ref{eq:selfsim}) with equality, so they contain 
enough information to determine the asymptotic threshold. They can also 
be used to determine a bound on the number of ``bad'' fault paths, as 
discussed in Section~\ref{sec:defs}, by considering only the low-order 
terms in the flow maps:
\begin{align}
\Gamma_w^1(\vec{\gamma}) & \leq 3\gamma_w^2 + 3\gamma_v^2 + 6\gamma_w\gamma_v \\
\Gamma_v^1(\vec{\gamma}) & \leq 3\gamma_v^2 + 16\gamma_v^3  = (3 + 16\gamma_v)\gamma_v^2.
\end{align}
These bounds clarify the relative contribution each component makes to the
failure probability of a fault-tolerant component. They also suggest a
conservative bound of $1/12$ on $\gamma_{th}$. This can be calculated 
by assuming that $\gamma_v=\gamma_w$, solving for the fixed point 
of the right hand side of each inequality, and taking the least such
fixed point.

\subsection{TRIPs for the $[3,1,3]$ code}
\noindent

What is the behavior of the wire and voter failure probabilities as the 
concatenation level $L$ increases? 
TRIPs based on Eqs~(\ref{eqn:wire})--(\ref{eqn:voter}) provide a visualization of each level crossing point for the two types of locations.  

\begin{figure}
\vspace*{13pt}
\centerline{\psfig{file=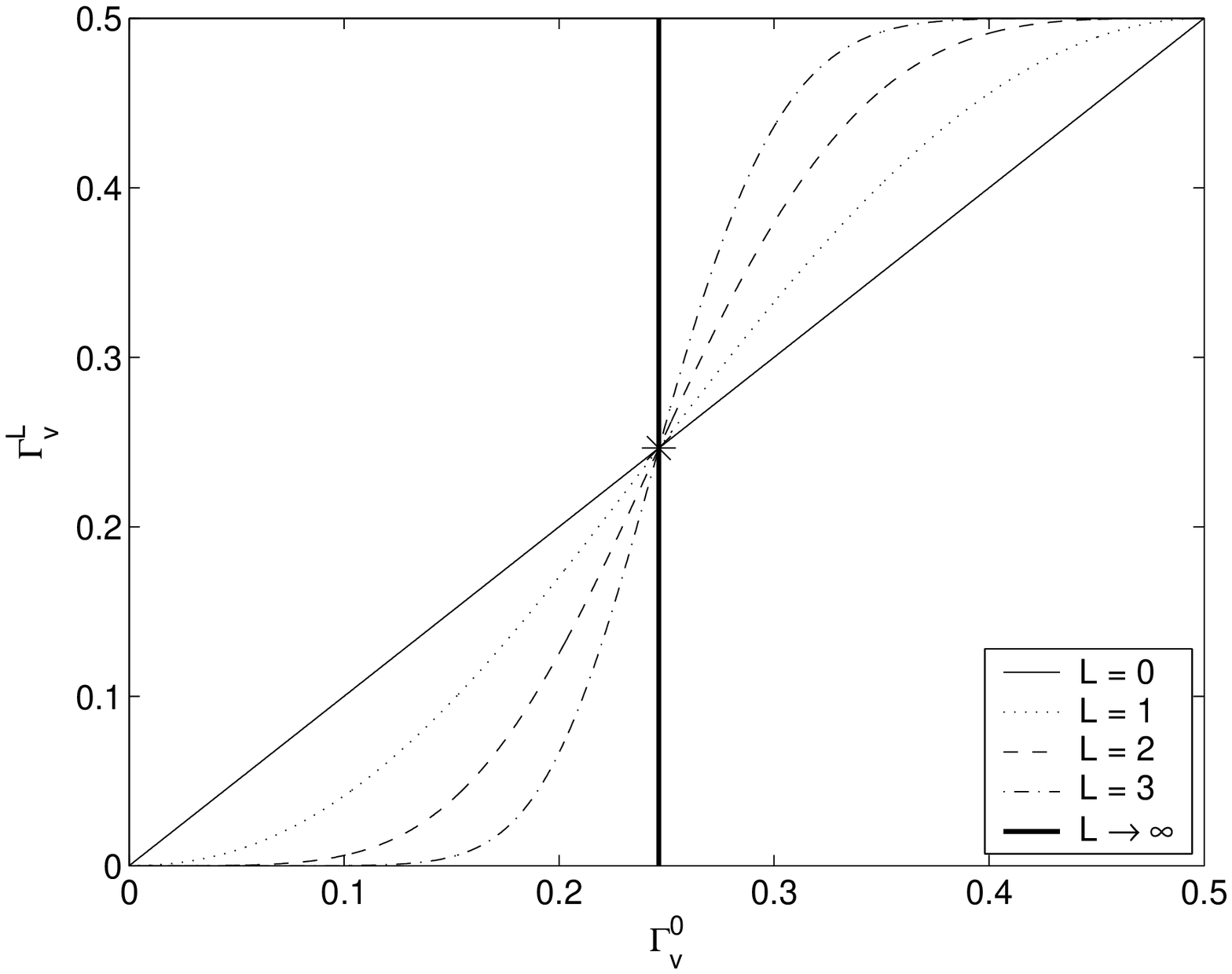,width=8.2cm}}
\vspace*{13pt} \fcaption{TRIP for a voter location for the $[3,1,3]$
code for $L=0,1,2,3,\infty$. Because $\Gamma_v^L$ is a function of only
$\gamma_v^0$ constructed by recursive application of $R(v)$, all of
the curves $\Gamma_v^L$ intersect at the same point. This point is
the fixed-point $\gamma_{th}\approx 0.246$ of the map and is
indicated by an asterisk.} \label{fig:votercurves}
\end{figure}

\begin{figure}
\vspace*{13pt}
\centerline{\psfig{file=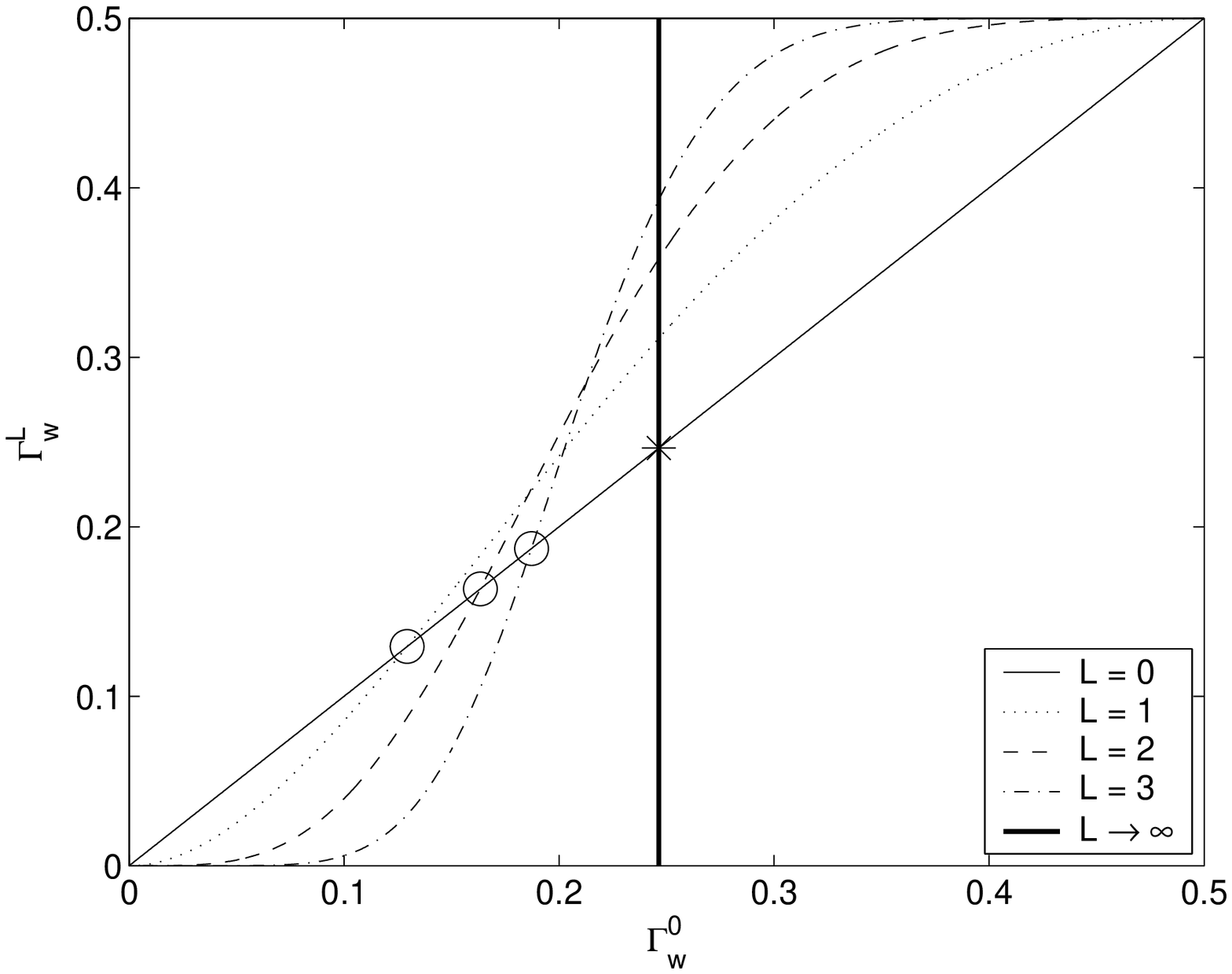,width=8.2cm}}\vspace*{13pt}
\fcaption{TRIP for a wire location for the $[3,1,3]$ code for
$L=0,1,2,3,\infty$ using the diagonal setting
$\gamma_w^0=\gamma_v^0$. Because $\Gamma_w^L$ is a function of both
$\gamma_w^0$ and $\gamma_v^0$ constructed by recursive composition,
the curves $\Gamma_w^L$ cross the $L=0$ line at different points.
Each of these points is a level-$L$ pseudothreshold indicated by a
circle, and the sequence of pseudothresholds $\gamma_w^1$,
$\gamma_w^2$, \dots, converges to the fault-tolerance threshold
$\gamma_{th}\approx 0.246$ indicated by an asterisk.}
\label{fig:wirecurves}
\end{figure}

Figures \ref{fig:votercurves} and \ref{fig:wirecurves} 
are TRIPs for the voter and wire locations, respectively. 
From 
Figure~\ref{fig:votercurves}, it is clear the voter probability 
$\Gamma_v^1(\vec{\gamma})$ is a one-parameter map, and thus should 
resemble the TRIP for the fault-tolerance threshold equation 
(Figure~\ref{fig:trp:a}). Since $\Gamma_v^1(\vec{\gamma})$ is a function of only 
$\gamma_v^0$ for all $L$, each $\Gamma_v^L(\vec{\gamma})$ intersects at the 
same fixed-point of the map. This fixed point $\gamma_{th}$, indicated by
a thick vertical line, is the $L=\infty$ pseudothreshold and the asymptotic threshold. 
It occurs at approximately $0.246$. Note that the $L=\infty$ pseudothreshold for a
particular location and the asymptotic threshold are not always equal, but they
happen to be in this example.

Even for a classical setting, there is a difference between 
pseudothresholds and the asymptotic threshold $\gamma_{th}$.
Figure~\ref{fig:wirecurves} shows the TRIP for the wire location at levels 
$L=0,1,2,3,\infty$. Here, unlike in the TRIP for the voter location, pseudothresholds 
appear in addition to an asymptotic threshold. This is because the 
replacement $R(w)$ for a wire includes locations of type $w,v,$ and $f$, 
creating a two-parameter map (fanouts are noiseless) that exhibits changing 
behavior with each concatenation level $L$. The curves now cross the $L=0$ 
line at different points. Each of these crossing points in the TRIP is a 
level-$L$ pseudothreshold. As we repeatedly replace the wire using $R(w)$, 
the number of voter locations begins to dominate, so the crossing point 
approaches the voter threshold, which corresponds to the asymptotic threshold.
The difference between the asymptotic threshold $\gamma_{th} \approx 0.246$ and the 
level-1 wire location pseudothreshold $\gamma_{w}^1 \approx 0.129$
is $0.117$, or $\gamma_{th}\approx 1.9 \times \gamma_{w}^1$.

\subsection{TIFDs for the $[3,1,3]$ code}
\noindent
  
Given that pseudothresholds can be so different from $\gamma_{th}$, can $\gamma_{th}$ be determined 
from just one application of the flow map?
In Section~\ref{sec:intro}, it was suggested that a TIFD provides an informative 
view of the effect of recursive simulation. 
Although a TRIP provides a visualization of the asymptotic behavior, it hides the fact that 
$\Gamma^1$ acts on a multidimensional space. When $\Gamma^1$ is chosen
such that Eq~(\ref{eq:selfsim}) is an equality, $\Gamma^1$ contains all of the information
about the flow since $\Gamma^L$ can be expressed in terms of $\Gamma^1$.
A TIFD shows where each point flows under repeated application of $\Gamma^1$ 
without iterating the map explicitly. 

\begin{figure}
\vspace*{13pt}\centerline{
\psfig{file=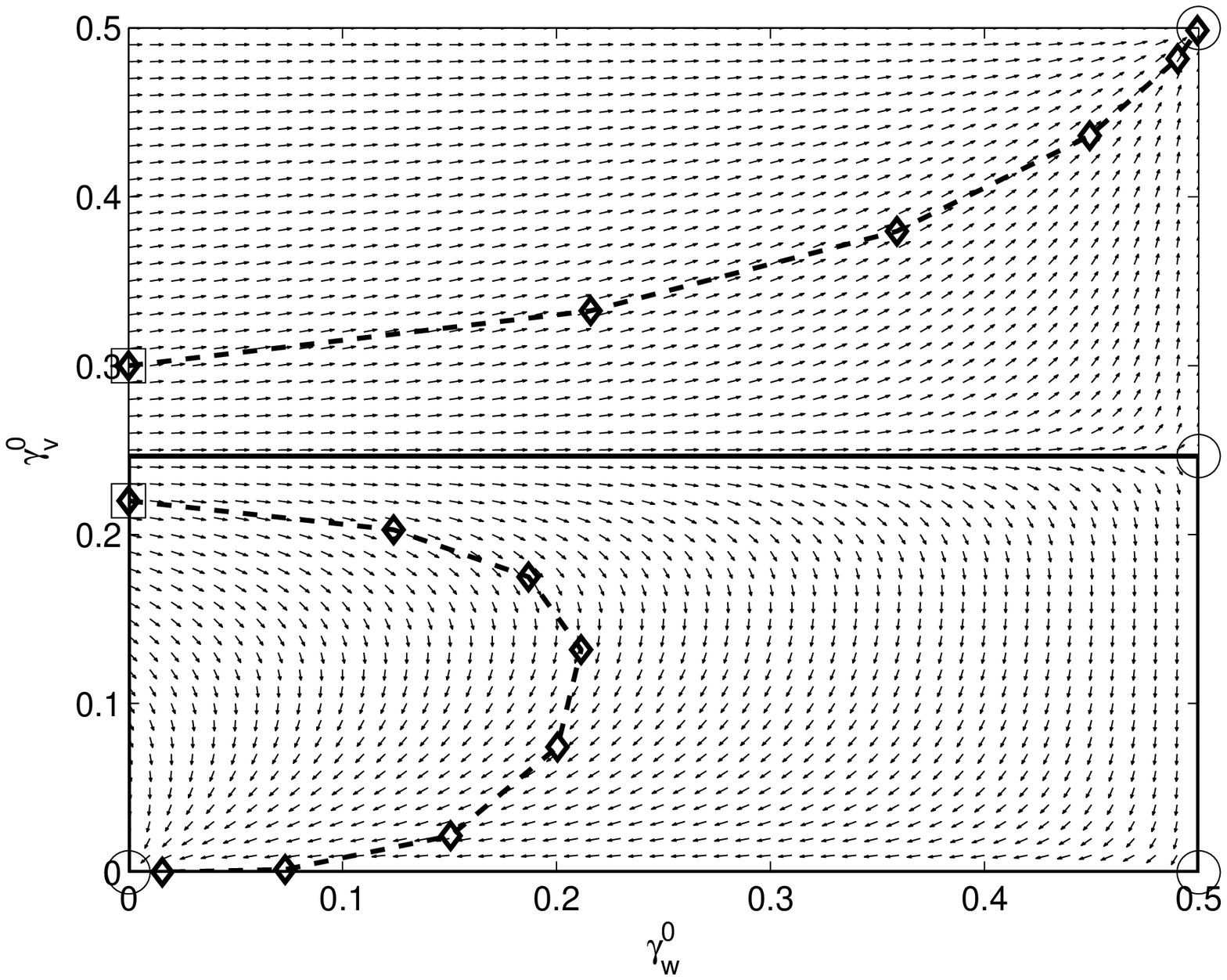,width=8.2cm}}\vspace*{13pt}
\fcaption{TIFD for $\gamma_w$ and $\gamma_v$ for the $[3,1,3]$ code.
The fixed points are indicated by circles. The region $[0,1/2)\times
[0,\gamma_{th})$ enclosed by a thick line is the set of points below
threshold, where $\gamma_{th}\approx 0.246$. Two sample trajectories
begin at the squares and flow along the thick dashed lines, where
the diamonds indicate the sequence of points as $L$ increases,
toward $(0,0)$ and $(1/2,1/2)$. } \label{fig:tmrvectorfield}
\end{figure}

In Figure~\ref{fig:tmrvectorfield}, a TIFD for wire and voter
failure probabilities on the unit half square is shown. 
The arrows represent the vectors 
$\Gamma^1(\vec{\gamma}) - \vec{\gamma}$, which give the probability flow under 
recursive simulation. There are five fixed points of the map: $(0,0)$, 
$(1/2,0)$, $(1/2,\gamma_{th})$, $(1/2,1/2)$, and $(1,0)$, where 
$\gamma_{th}\approx 0.246$ cannot be expressed in radicals. Circles mark four
of these fixed points. The subthreshold 
region $T$ is $[0,1/2)\times [0,\gamma_{th})$, indicated by the thick black 
box. Three corners of the subthreshold region are fixed points of the map. 
The fault-tolerance threshold $\gamma_{th}$ is the size of the largest 
``cube'', a square in this case, that is contained in the subthreshold 
region.

First, if $\gamma_w^0=0$ and $\gamma_v^0>0$, then the flow draws these points
off of the $\gamma_v^0$--axis. This occurs because the wire replacement
rule $R(w)$ contains both voter and wire locations, so the failure
probabilities ``mix''. Second, if $\gamma_v^0=0$, then any point
$\gamma_w^0<1/2$ flows to the origin because the voters amplify any bias
toward success. Third, the voter failure probability $\Gamma_v^1(\vec{\gamma})$ is
independent of $\gamma_w^0$, so the voter probability is a simple
one-parameter map under replacement. If $\gamma_v^0<\gamma_{th}$,
the map's fixed point, then the voter probability flows toward the
$\gamma_w^0$--axis. Finally, if $\gamma_v^0<\gamma_{th}$ and $\gamma_w^0<1/2$,
then initially the wire failure probability may increase because the
voters are not reliably correcting errors. However, the voters improve
with each iteration, so eventually this trend reverses. The voters
begin correcting more errors than they introduce, and all of these
points flow toward the origin.

For the classical $[3,1,3]$ code, the TIFD fully characterizes the threshold 
set $T$ for three reasons. First, deviated inputs, i.e., inputs that are not 
codewords, are corrected before faults are introduced by the simulated gate 
locations. Second, the flow maps are the precise component failure 
probabilities. Third, there is no phase noise in classical fault-tolerance,
so the parameters of the noise channel only change in the trivial way.
Furthermore, the entire flow is easily visualized since the TIFD is 
two-dimensional. Under these conditions, the TIFD is an ideal tool for 
understanding and visualizing the process by which recursive simulation 
improves reliability and exhibits a threshold.

\section{Quantum Pseudothresholds}
\label{sec:quantum}
\noindent

We have seen that pseudothresholds exist even in a simple classical fault-tolerance
scheme. In a quantum fault-tolerance scheme, the tools we have developed can now 
be applied to determine sequences of quantum pseudothresholds. In this section, 
we study thresholds for quantum fault-tolerance using the $[[7,1,3]]$ CSS code.  
We follow the circuit construction given in \cite{steane02overhead}. 
As in the classical example, TRIPs are again used to identify pseudothresholds 
for the given location types using particular settings, allowing us to determine 
the reliability achieved with each level of concatenation.  In addition, we 
characterize the failure probability map using TIFDs. 

\subsection{The $[[7,1,3]]$ code and its flow map}
\noindent

The $[[7,1,3]]$ quantum code encodes a single qubit in 7 qubits 
with distance 3, meaning it corrects a general quantum error on a
single qubit. The set of location types $\myell$ involved in the error 
correction routine is $\Omega = \{1,2,w,1m,p\}$:
\begin{itemize}
\item $\myell=1$: one-qubit gate
\item $\myell=2$: two-qubit gate
\item $\myell=w$: wait (memory) location
\item $\myell=1m$: one-qubit gate followed by a measurement \cite{svore04local}, since the replacement rule for a measurement contains no error correction and since measurements in the networks for the 7-qubit code only appear after one-qubit gates 
\item $\myell=p$: ancilla preparation location, which we model as a one-qubit 
gate for simplicity \cite{svore04local}
\end{itemize}

We consider the depolarizing error model, where a location $\myell$ fails independently with 
probability $\gamma_{\myell}$.
In our probabilistic error model, we assume for a location 
failure $\gamma_\myell$ on a single qubit, a $X$,$Y$, or $Z$ error occurs with 
probability $\gamma_\myell/3$. The distinction between $X$,$Y$, and $Z$ errors 
is used in the approximation analysis.

\begin{figure}[htbp]
\vspace*{13pt}\centerline{\psfig{file=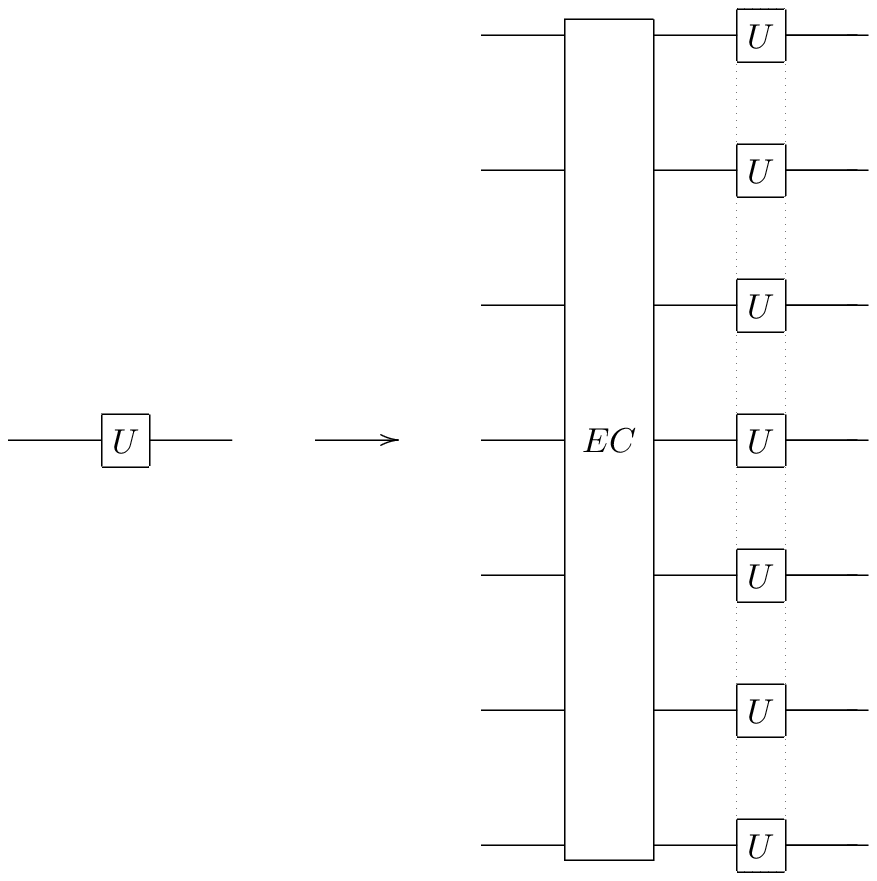,width=8.2cm}}\vspace*{13pt}
\fcaption{Replacement rule for a one-qubit gate $U$. The replacement
includes error correction ($EC$) followed by a fault-tolerant
implementation of the one-qubit gate $U$, indicated by a dashed
box.} \label{fig:1replace}
\end{figure}

As in the classical case, we define a replacement rule for each type of location. We 
use the same replacement rules as given in \cite{svore04local}, where we replace 
each location by error correction followed by a fault-tolerant implementation of 
the location. For example, a one-qubit gate is replaced by error correction and 
a transversal one-qubit gate, as shown in Figure \ref{fig:1replace}. 

We use the same approach to derive the failure probability map as in the classical example, 
with the caveat that we do approximate counting of the ways in which two or more errors occur 
on the data qubits during the execution of the concatenated circuit.
The composition of this map approximates the behavior of the concatenated circuit.
This means that threshold results derived from this map are also approximate. 
The map does not account for incoming errors since when failure probabilities 
are below threshold, the probability of incoming errors should typically be 
small, and thus should not greatly affect the probability of two or more errors 
on the data qubits.
The details of the failure probability map are given in 
Section IV B of \cite{svore04local}.

\subsection{TRIPs for the $[[7,1,3]]$ code}
\noindent

\begin{figure}
\vspace*{13pt}\centerline{
\psfig{file=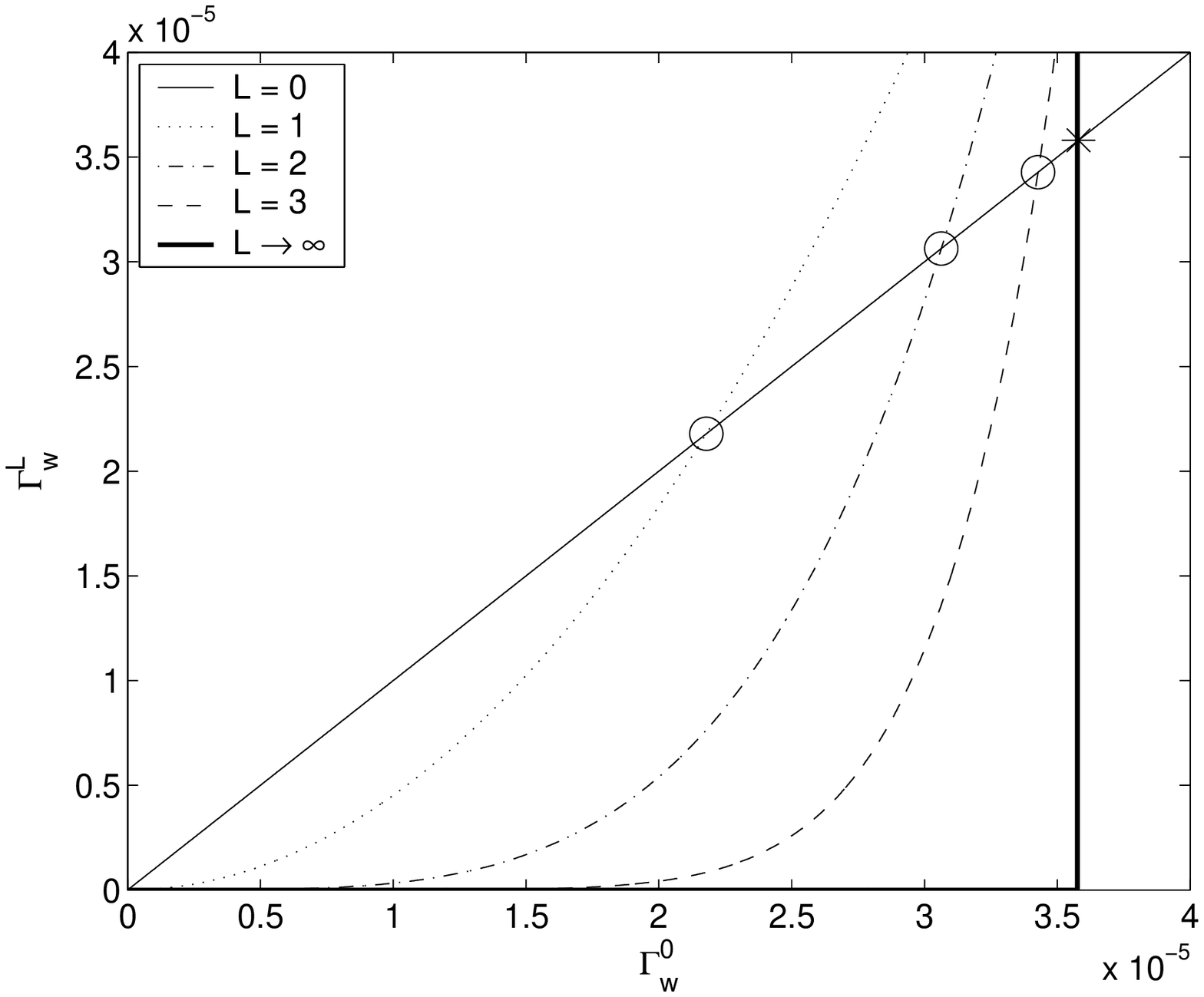,width=8.2cm}}\vspace*{13pt}\fcaption{TRIP
for a wait location for $L=0,1,2,3,\infty$ for the initial setting
$\gamma_w^0 = 1/10(\gamma_i^0)$, where $i$ indicates all location
types except for a wait location. Circles indicate pseudothresholds
and an asterisk marks the threshold for this gate and setting. 
The pseudothresholds occur at probabilities $2.2\times 10^{-5}$, 
$3.0\times 10^{-5}$, and $3.4\times 10^{-5}$. The wait threshold
for this setting is $3.6\times 10^{-5}$. This corresponds to
$\gamma=3.6\times 10^{-4}$ in the Steane setting.
} \label{fig:qw}
\end{figure}

\begin{figure}
\vspace*{13pt}\centerline{
\psfig{file=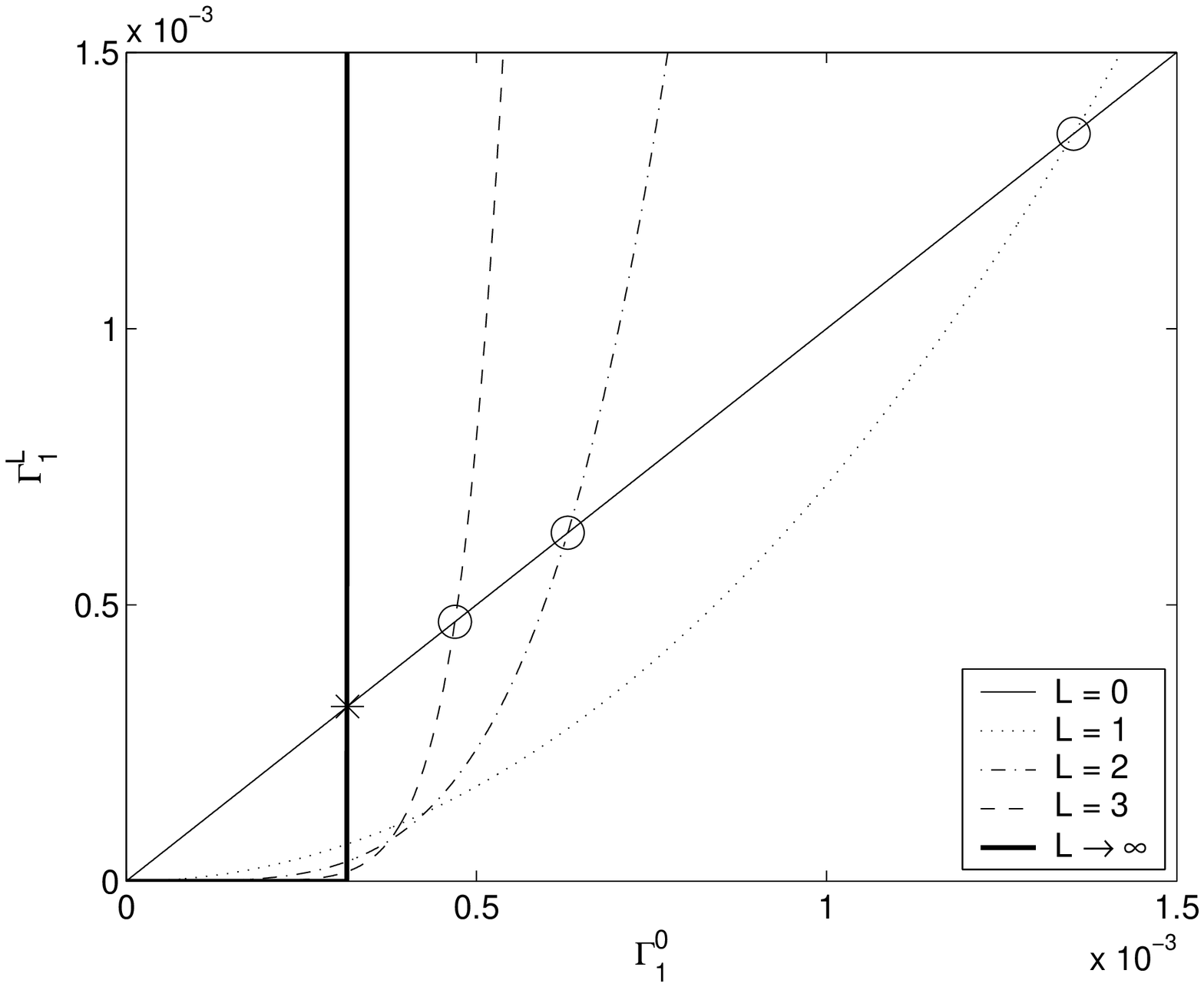,width=8.2cm}}\vspace*{13pt}
\fcaption{TRIP for a one-qubit gate location for $L=0,1,2,3,\infty$
for the initial setting $\gamma_w^0 = 1/10(\gamma_i^0)$, where $i$
indicates all location types except for a wait location. Circles
indicate pseudothresholds and an asterisk marks the threshold for
this gate and setting. The pseudothresholds occur at probabilities
$1.4\times 10^{-3}$, $6\times 10^{-4}$, and $5\times 10^{-4}$. The
one-qubit gate threshold for this setting is $3\times 10^{-4}$.
} \label{fig:q1}
\end{figure}

\begin{figure}
\vspace*{13pt}\centerline{
\psfig{file=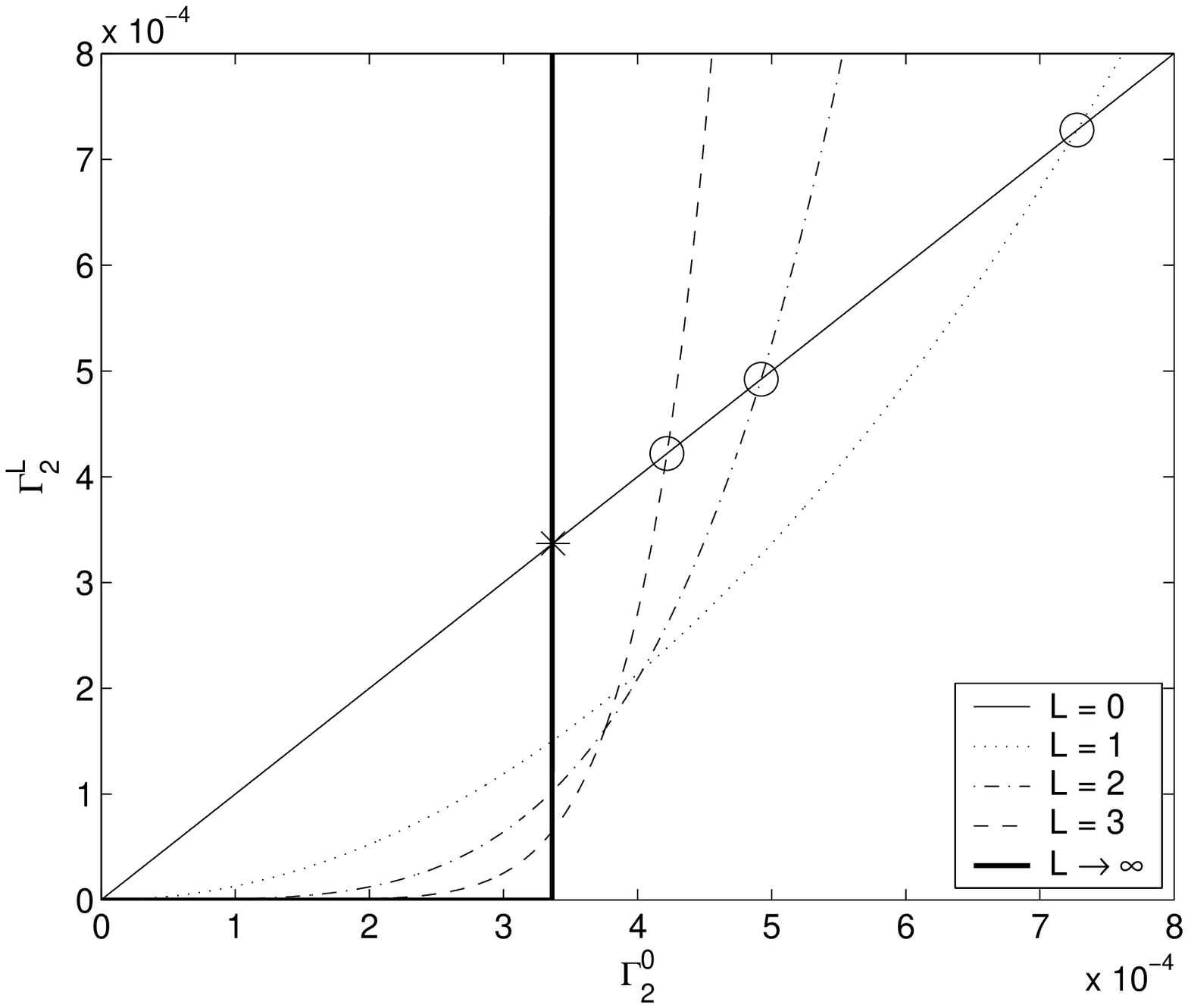,width=8.2cm}}\vspace*{13pt}
\fcaption{TRIP for a two-qubit gate location for $L=0,1,2,3,\infty$
for the initial setting $\gamma_w^0 = 1/10(\gamma_i^0)$, where $i$
indicates all location types except for a wait location. Circles
indicate pseudothresholds and an asterisk marks the threshold for
this gate and setting. The pseudothresholds occur at probabilities
$7.3\times 10^{-4}$, $4.9\times 10^{-4}$, and $4.2\times 10^{-4}$.
The two-qubit gate threshold for this setting is $3.4\times 10^{-4}$.
}  \label{fig:q2}
\end{figure}

To determine pseudothresholds, we plot the reliability of each 
component at each level of concatenation using the flow maps.
Figures~\ref{fig:qw}--\ref{fig:q2} show TRIPs for wait, one-qubit, and 
two-qubit locations for the Steane setting 
$g(\gamma) = (\gamma,\gamma,\ldots,\gamma,\gamma/10)$, where the last 
component is the wait location failure probability. 

These results in this section are obtained using a flow map, but they 
are verified using a Monte-Carlo simulation for one level of code 
concatenation. The Monte-Carlo simulation method randomly generates faults 
within the fault-tolerant circuit representing the original location according
to the failure probabilities $\gamma_\myell$. These faults create errors 
that propagate to the output of the circuit. A particular set of faults 
may or may not create too many errors at the circuit's output. By running
on the order of one million simulations, we can estimate the failure
probability of the original location. The particular Monte-Carlo 
simulation includes the ancilla preparation networks but does not model 
input errors so as to agree with the assumptions under which the flow map 
is derived.

Figure~\ref{fig:qw} shows the TRIP for a wait location at 
levels $L=0,1,2,3,\infty$. 
Since the flow map is a multi-parameter map, 
the crossing points no longer cross the line $L=0$ at the same point and 
thus pseudothresholds appear at each level. 
As $L$ increases, the level-$L$ 
pseudothreshold approaches an asymptotic threshold that depends on the
location and setting. Note that the region between the $L=1$ curve and the 
$L=0$ line is quite small for these initial conditions, smaller than 
the region considered to be below the asymptotic threshold. This is similar 
to the behavior of the classical wait location. The behavior is largely due
to the Steane setting -- the level-1 simulation of the wait location includes
other location types that have been set to fail with probability $10 \times \gamma_w^0$.
However, as the level of concatenation of the wait location increases, the trade-offs 
between the failure probabilities of the location types begins to stabilize causing the 
level-4 pseudothreshold, for example, to be much larger and closer to the asymptotic 
threshold. The level-$1$ pseudothreshold as calculated from the flow map
is $2.2\times 10^{-5}$ and the corresponding value from Monte-Carlo simulation
is approximately $2.4\times 10^{-5}$.

Figure~\ref{fig:q1} shows the TRIP for the one-qubit location type for levels 
$L=0,1,2,3,\infty$.  The level-1 pseudothreshold is about $4.6$ times greater 
than the asymptotic one-qubit gate threshold since the replacement $R(1)$ 
includes many wait locations, whose initial setting is one-tenth of the 
initial one-qubit gate failure probability. The Monte-Carlo simulation
gives a level-$1$ pseudothreshold of approximately $1.3\times 10^{-3}$ versus
an estimate of $1.4\times 10^{-3}$ from the flow map.

Similarly, for a two-qubit gate location, the level-1 pseudothreshold is a
factor of $2$ larger than the asymptotic two-qubit gate threshold 
(Figure \ref{fig:q2}). Note that 
the level-1 two-qubit gate pseudothreshold is about half the size of the level-1 one-qubit 
gate pseudothreshold. This is because error correction is required on two logical 
qubits and thus there are twice the number of locations in a one-qubit gate 
replacement. The level-$1$ pseudothreshold here is about $7.3\times 10^{-4}$
versus a Monte-Carlo estimate of about $6.4\times 10^{-4}$.

From the threshold reliability information plots (TRIPs), it is apparent that 
multi-parameter maps and higher levels of concatenation are required to 
determine a threshold result. Across location types, the largest level-$1$
pseudothreshold is approximately $40$ times larger than the smallest
asymptotic gate threshold. The smallest asymptotic gate threshold is
the memory threshold, so it is appropriate to scale this gate threshold
by $10$ to eliminate artifacts from the setting. The pseudothreshold-threshold
factor then becomes $4.6$ for this example.

\subsection{TIFDs for the $[[7,1,3]]$ code}
\noindent

We use the TIFD to characterize the flow of the maps based on the 
semi-analytical methods of \cite{svore04local}.  By using a TIFD instead of the 
TRIP, the flow of the failure probabilities as well as pseudothresholds can be 
visualized. In the quantum case, however, the TIFD is a 4-dimensional 
flow that is challenging to visualize.  Instead, we take 2-dimensional 
projections to determine the flow.

\begin{figure}
\vspace*{13pt}\centerline{
\psfig{file=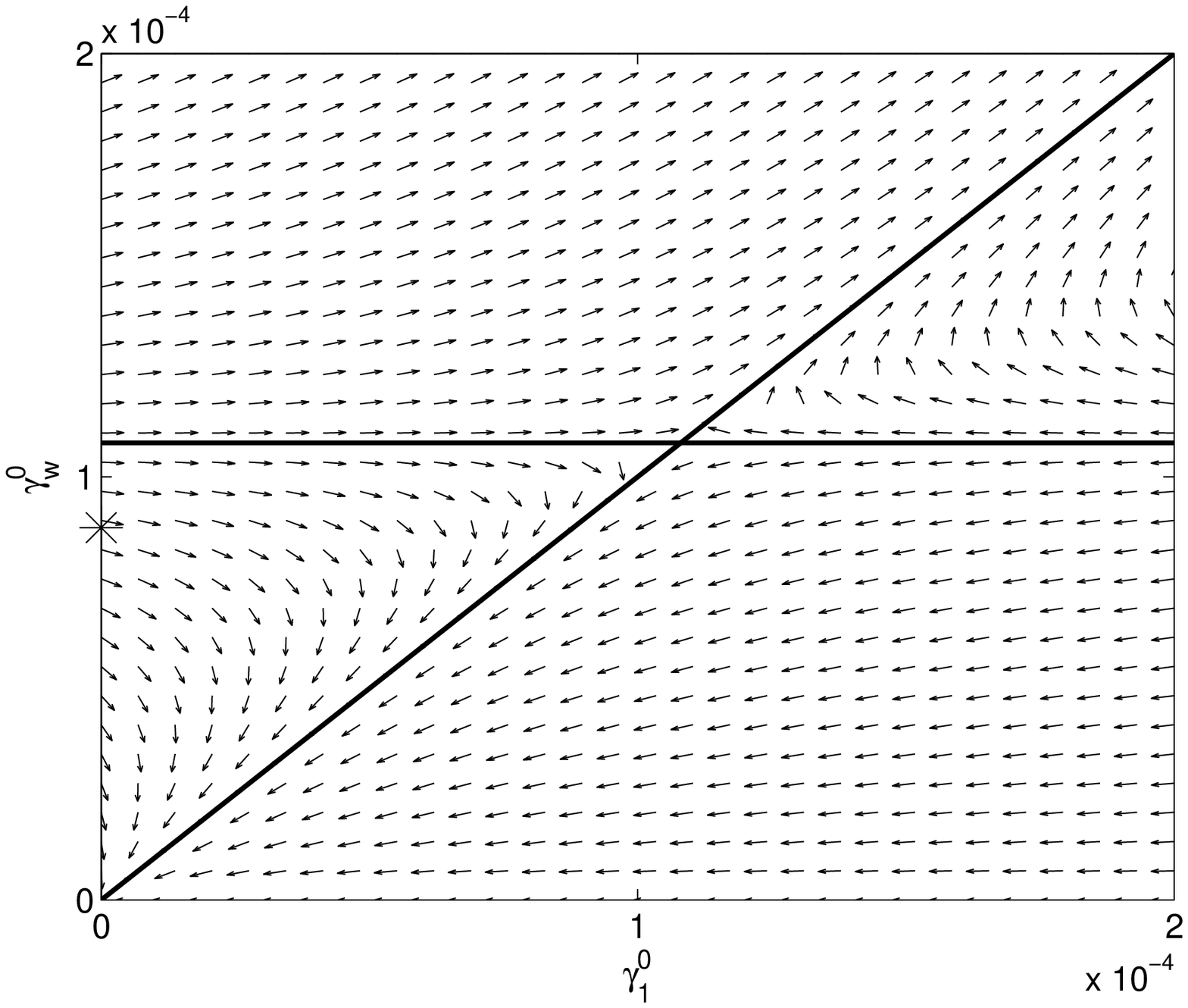,width=8.2cm}}\vspace*{13pt} \fcaption{TIFD
projected onto the $\gamma_1$--$\gamma_w$ plane, where $\gamma_l^0 =
0$ for all $l$ except $\gamma_w^0$ and $\gamma_1^0$. The arrows
represent the vector field flow $\Gamma^1(\vec{\gamma}) -
\vec{\gamma}$.  The thick lines illustrate the
separatrices. The horizontal separatrix appears to have zero slope, but
it intersects the $\gamma_1^0$-axis at about $10^{-2}$. The asterisk 
marks the wait gate asymptotic threshold for the setting
in which $\gamma_w=\gamma$ and $\gamma_i=0$ for all other location types.}
\label{fig:qflow1w}
\end{figure}

\begin{figure}
\vspace*{13pt}\centerline{
\psfig{file=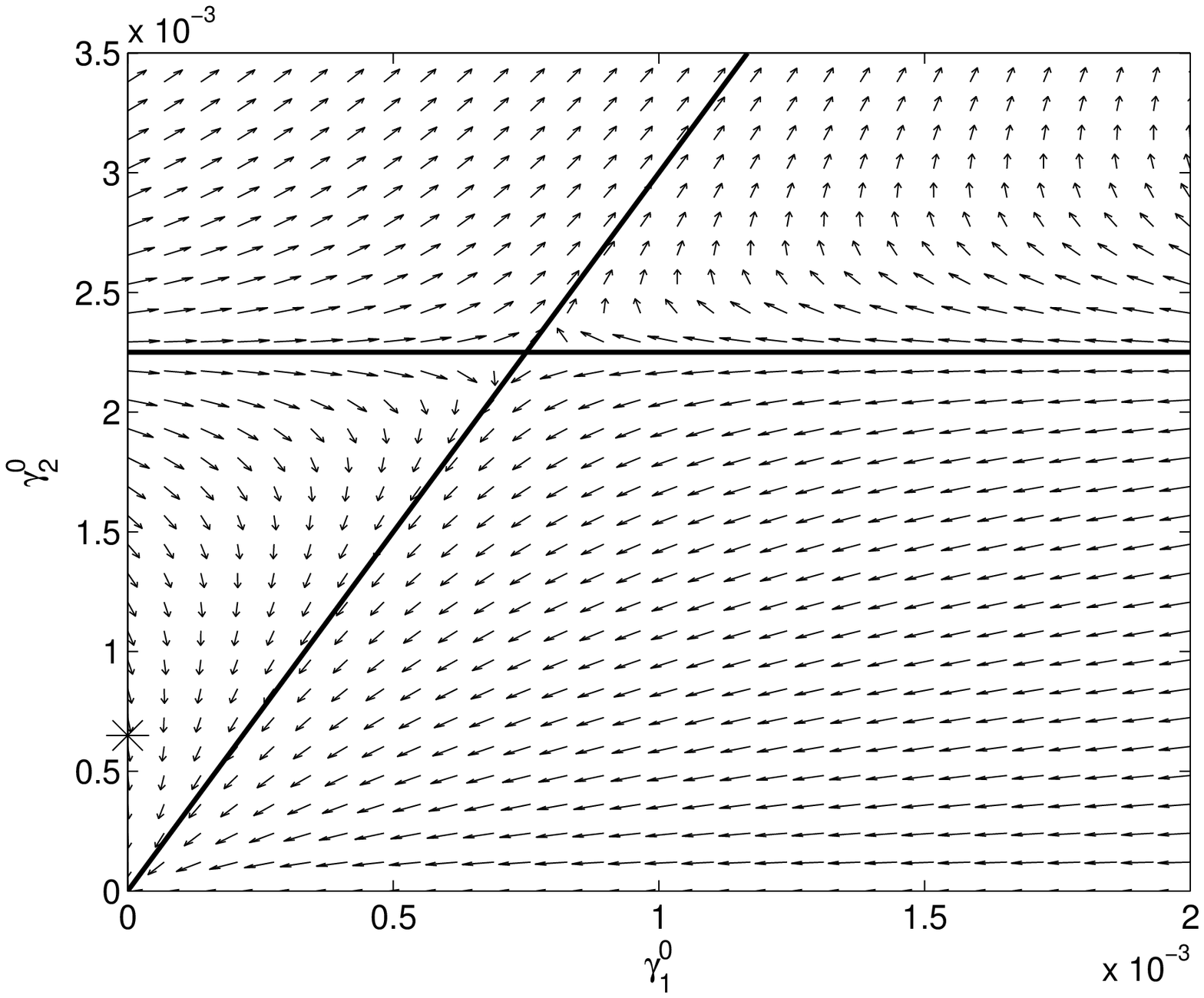,width=8.2cm}}\vspace*{13pt} \fcaption{TIFD
projected onto the $\gamma_1$--$\gamma_2$ plane, where $\gamma_l^0 =
0$ for all $l$ except $\gamma_1^0$ and $\gamma_2^0$. The arrows
represent the vector field flow $\Gamma^1(\vec{\gamma}) -
\vec{\gamma}$.  The thick lines illustrate the
separatrices. The horizontal separatrix appears to have zero slope,
but it intersects the $\gamma_1^0$-axis at about $10^{-2}$. We do
not show this intersection because the intersection with the
$\gamma_2^0$-axis at about $2.3\times 10^{-3}$ has a more
significant role in determining the threshold. The asterisk marks
the two-qubit gate threshold for the setting in which $\gamma_2=\gamma$
and $\gamma_i=0$ for all other location types.} \label{fig:qflow12}
\end{figure}

Figures~\ref{fig:qflow1w}--\ref{fig:qflow12} show TIFDs involving location types 
$l=1,2,w$. Figure \ref{fig:qflow1w} shows the vector field 
$\Gamma^1(\vec{\gamma}) - \vec{\gamma}$ projected onto the 
$\gamma_1$--$\gamma_w$ plane.  Note that the flows are partitioned by separatrices
shown by the thick black lines. The map is independent 
of concatenation level in our approximation, so this flow fully characterizes the 
behavior of the map. The $\gamma_w$ threshold found by the horizontal separatrix
appears around $1.1\times 10^{-4}$. The asymptotic threshold for the map restricted 
to the $\gamma_w$--axis is indicated by an asterisk on the $\gamma_w$--axis. 
Note the separatrix indicates a wait location pseudothreshold.

Figure~\ref{fig:qflow12} is the TIFD projected onto the 
$\gamma_1$--$\gamma_2$ plane. Again, the flows form a separatrix around 
$\gamma_2=2 \times \gamma_1$.  This is because there are two error correction 
routines in a two-qubit gate replacement, compared to only one error 
correction routine in the replacement for a one-qubit gate.  The $\gamma_2$ 
threshold appears along the other separatrix around $2.3 \times 10^{-3}$.
The asymptotic threshold restricted to the $\gamma_2$--axis, indicated by the 
asterisk, is a factor of $3.8$ below the pseudothreshold.

In the classical setting, the TIFD indicated the asymptotic threshold, since the map was exact and only two-dimensional. 
However, it is evident from the TIFDs for the $[[7,1,3]]$ code that 2-dimensional projections of the flow are 
insufficient to determine the quantum fault-tolerance threshold set $T$. Since the threshold set is a multi-dimensional 
surface, the two-dimensional projection fails to indicate flow in the other dimensions. Although it appears the 
separatrices indicate a separation between points that flow to zero and those that flow to one, it cannot be used to 
determine the threshold, but it may be used to determine an upper bound on the threshold for this example.

\section{Techniques for Determining the Asymptotic Threshold}
\label{sec:technique}
\noindent

In Section~\ref{sec:quantum}, low-dimensional projections of the flow using a TIFD were used 
to establish pseudothresholds.  However, the TIFD could not be used to determine the 
asymptotic threshold $\gamma_{th}$. It may be possible, though, to bound $\gamma_{th}$ for a particular map by 
restricting the map $\Gamma$ to the axes. In this section, we describe a possible 
technique for upper bounding the fault-tolerance threshold $\gamma_{th}$.

Consider the following setting, the {\it axis setting}, where every initial failure 
probability is $0$, except for the axis of interest, i.e., $g(\gamma) = (\gamma,0,0,\ldots,0)$, and 
$\gamma$ is assigned to the location axis of interest. We conjecture that the level-$1$
pseudothreshold for this setting upper bounds $\gamma_{th}$.

\begin{figure}
\vspace*{13pt}\centerline{
\psfig{file=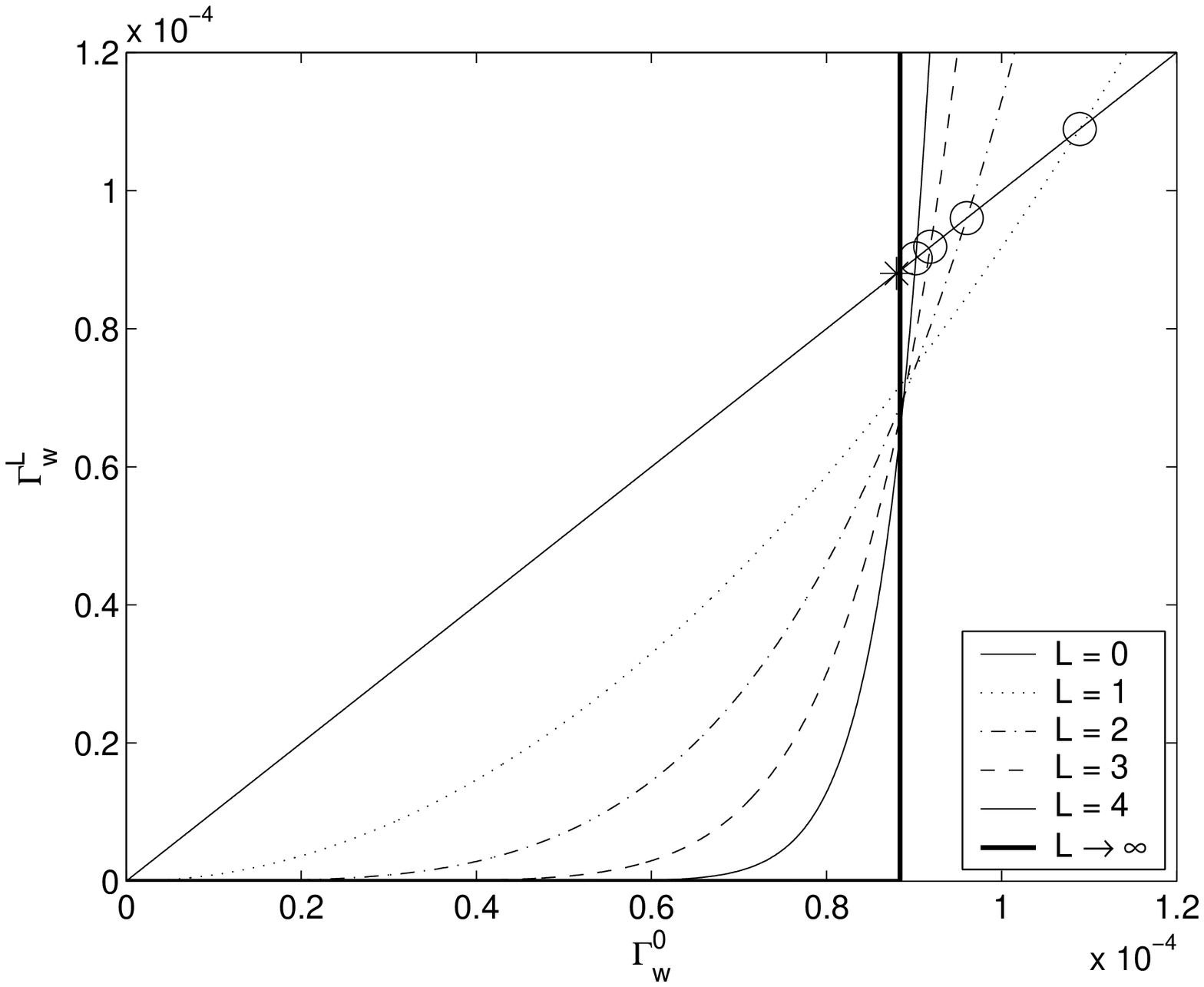,width=8.2cm}}\vspace*{13pt}\fcaption{TRIP
for a wait location for $L=0,1,2,3,4,\infty$ for the axis setting. 
The pseudothresholds are $1.1\times 10^{-4}$, $9.6\times 10^{-5}$,
$9.2\times 10^{-5}$, and $9.0\times 10^{-5}$. The threshold for
the wait location for this setting is $8.8\times 10^{-5}$.
}
\label{fig:SCCAgw}
\end{figure}

Consider the plot shown in Figure~\ref{fig:SCCAgw} of the approximated 
$\gamma_w$ pseudothresholds of the $[[7,1,3]]$ code for the axis setting. Note that the 
pseudothresholds for the wait location for the axis setting are strictly 
decreasing toward a threshold. We find similar behavior for the other location types as well.
These pseudothresholds and the threshold found for the wait location 
are lower than the pseudothresholds and thresholds for the other location types 
in the axis setting. This suggests that the $L=1$ wait location pseudothreshold for the 
axis setting, the smallest level-$1$ axis pseudothreshold, is an upper bound on $\gamma_{th}$.

\begin{figure}
\vspace*{13pt}\centerline{
\psfig{file=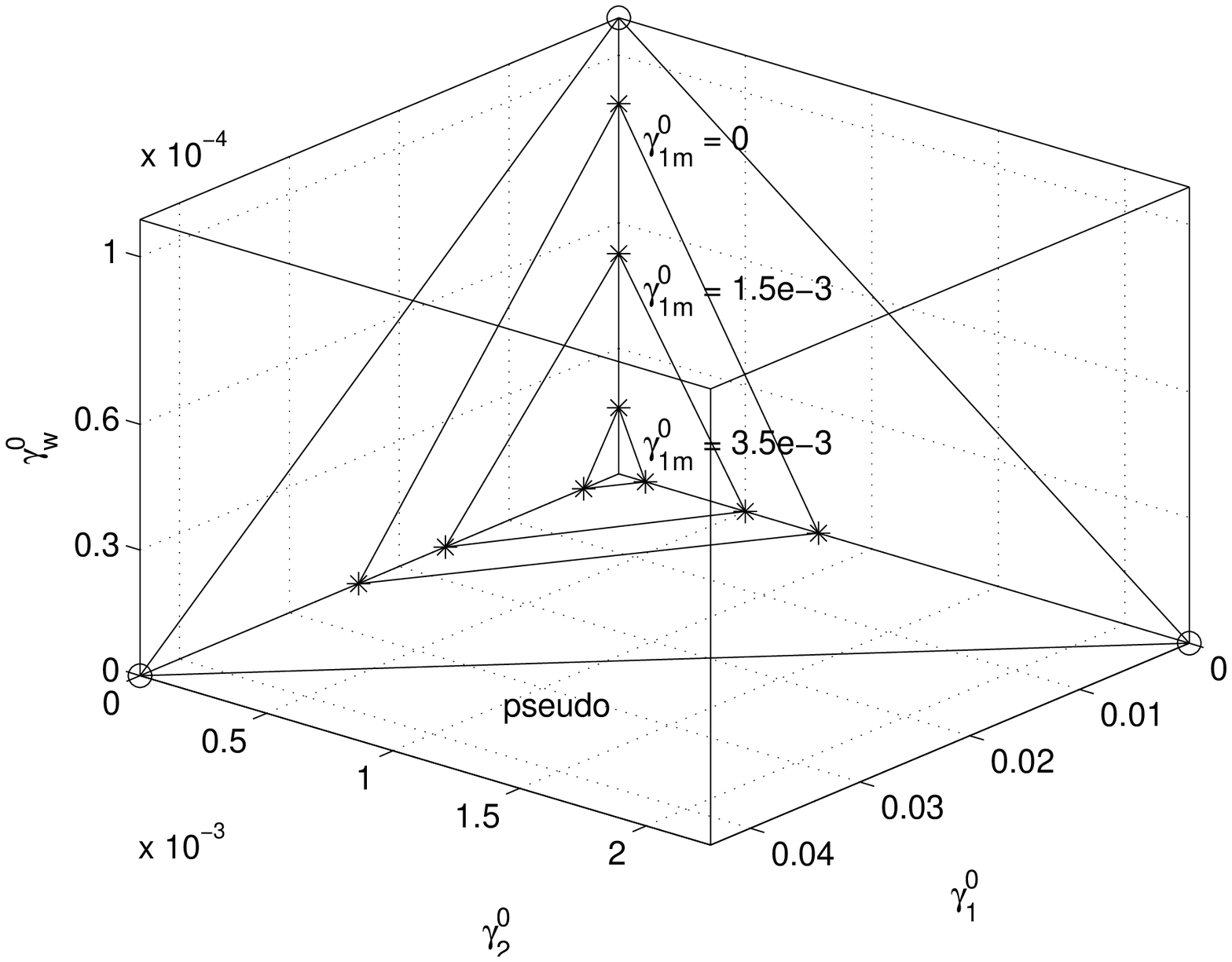,width=8.2cm}}\vspace*{13pt} \fcaption{The set
$T$ for the $[[7,1,3]]$ code together with a slice of the convex
hull of the level-$1$ axis pseudothresholds. The level-$1$ axis
pseudothresholds are plotted with open circles and connected with
lines to illustrate their convex hull. The other three hulls are
schematic representations of the numerically computed boundary of
the set $T$ for varying values of $\gamma_{1m}^0$. All points
beneath a given hull are below threshold. The largest cube contained
in $T$ has edge length $\gamma_{th}\approx 8.8\times 10^{-5}$. 
Note that this does not contradict the wait location threshold in
Figure~\ref{fig:qw} because the other locations are $10$ times
less reliable in that calculation.}
\label{fig:basin}
\end{figure}

This conclusion that pseudothresholds with the axis setting provide an upper bound on the fault-tolerance threshold is supported by numerical evaluation of the threshold set $T$ in 
Figure~\ref{fig:basin}. This figure shows four convex hulls: one pseudothreshold hull and three threshold hulls.
The pseudothreshold hull is determined by the $\gamma_1^1$, $\gamma_2^1$, and $\gamma_w^1$ axis pseudothresholds,
which are plotted as circles, while the threshold hulls were determined numerically from the flow map for a
grid of parameters. The region of parameter space above the pseudothreshold hull is strictly above threshold for any value of $\gamma_{1m}$.
The largest threshold hull corresponds to $\gamma_{1m}^0=0$ and all points beneath it are below threshold.
Similarly, the other two threshold hulls correspond to $\gamma_{1m}^0=1.5\times 10^{-3}$ and 
$\gamma_{1m}^0=3.5\times 10^{-3}$.

Interestingly, as our choice of language indicates, $T$ appears to be a convex set equal to the convex hull of the axes
thresholds and the origin. The edge length of the largest cube in $T$ is approximately $\gamma_{th}\approx 8.8\times 10^{-5}$.
Furthermore, $T$ appears to be contained in the convex hull of the axes pseudothresholds and the origin. These pseudothresholds
are $\gamma_1^1\approx 4.4\times 10^{-2}$, $\gamma_2^1\approx 2.3\times 10^{-3}$, and $\gamma_w^1\approx 1.1\times 10^{-4}$ for the corresponding axes
settings.
The $\gamma_{1m}^1$ pseudothreshold is comparable to $\gamma_1^1$. The smallest level-$1$ pseudothreshold is 
$\gamma_w^1\approx 1.1\times 10^{-4}$, so this is an upper bound on $\gamma_{th}$ for this example. Though the error correction
networks are slightly different, this upper bound does not contradict a rigorous lower bound of $2.73\times 10^{-5}$ for the 
same code \cite{aliferis05}.

To further confirm the results found using the semi-analytical map, 
we use a Monte-Carlo simulation to determine 
pseudothresholds for the [[7,1,3]] code. 
We find the following pseudothresholds with the axis setting by fitting a 
quadratic to the numerical TRIPS: $\gamma_1^1\approx 8.0\times 10^{-2}$, 
$\gamma_2^1\approx 1.7\times 10^{-3}$, and $\gamma_w^1\approx 1.5\times 10^{-4}$. 
The $\gamma_1^1$ pseudothreshold differs by a factor of $1.81$ from the $\gamma_1^1$ pseudothreshold found using the flow map.
The $\gamma_2^1$ and $\gamma_w^1$ pseudothresholds found using 
the flow map differ by a factor of $0.73$ and a factor of $1.36$, respectively,  
from numerical calculations based on a Monte-Carlo simulation. 
These differences could be reduced by revisiting
some of the approximations in the flow map derivation.

While we do not prove our conjecture, we offer two supporting observations. The first is an observation that follows 
from the fact that the flow map has a threshold. There are positive integers $A$ and $t$ such that
$\Gamma_{\myell}^1(\vec{\gamma})\leq A\gamma_{max}^{t+1}$ for all $\myell$, 
where $\gamma_{max}\equiv \hbox{max}\ \gamma_{\myell}$.
These integers determine the well-known lower bound $A^{-1/(t+1)}\equiv \gamma_{th}^{min}\leq\gamma_{th}$
on the threshold.

The next observation is that if $\Gamma$ causes all components of $\vec{\gamma}$ to increase or remain 
unchanged, then $\vec{\gamma}$ is above the established lower bound $\gamma_{th}^{min}$. More precisely,
if $\Gamma_{\myell}^1(\vec{\gamma})\geq\gamma_{\myell}$ for all $\myell$, then $\gamma_{\myell}\geq\gamma_{th}^{min}$ 
for at least one $\myell$. This is true because if $\gamma_{\myell}<\gamma_{th}^{min}$ for all $\myell$, then 
$\gamma_{max}<\gamma_{th}^{min}$. In particular, 
$\Gamma_{\myell}^1((\gamma_{max},\gamma_{max},\dots,\gamma_{max}))\leq A\gamma_{max}^{t+1}<\gamma_{max}$.

A $\vec{\gamma}$ satisfying the second observation is not necessarily above 
threshold, but we conjecture that this second observation remains true
when $\gamma_{th}^{min}$ is replaced by $\gamma_{th}$ for maps $\Gamma$ 
describing failure probabilities under independent stochastic error models. 
We know this to be the case for one-dimensional maps. If the conjecture is true 
in general, then the level-$1$ pseudothreshold for the axis setting upper 
bounds $\gamma_{th}$ for a particular map.

\section{Conclusions and Future Work}\label{sec:conclusion}
\noindent

Pseudothresholds are the failure probabilities below which recursive simulation
improves the reliability of a particular component. Yet, pseudothresholds
have been shown to be up to a factor of $4$ greater than the asymptotic
threshold for the Steane setting and more than a factor of $10$ different
for the axes settings. This behavior is a generic phenomenon in both classical
and quantum fault-tolerance. The tools we have presented provide a way to
visualize pseudothresholds and thresholds, and we conjecture that for a
given setting, they may provide an upper bound on the fault-tolerance threshold
for a particular map.

Pseudothreshold behavior also influences the accuracy of some
quantum threshold estimates. If some of the reported threshold estimates are
actually pseudothresholds, the examples we have given suggest that these
estimates may be inaccurate by a factor of $4$ or more. These observations
apply, in particular, to some numerical threshold estimates.
However, other factors such as the noise model and circuit construction more 
greatly influence the threshold value and by making a judicious choice of 
concatenation level and setting, the difference between estimates and an 
asymptotic threshold can be reduced.

Although a fault-tolerance threshold for infinite scalability cannot be determined by low-level pseudothresholds, pseudothresholds will become important design parameters in engineering a quantum computer.  In practice, quantum computers may operate very close to threshold and require only a few levels of recursive simulation.  If this is the case, then pseudothresholds can help determine design trade-offs and the required relative reliability of circuit components.  In addition, the difference between pseudothresholds can be used to determine an appropriate level of concatenation that is within current physical capabilities.

It is important to note that pseudothresholds can also be used to determine the frequency of error correction required for certain location types.
If, for example, a level-1 pseudothreshold for a wait location shows the failure probability worsens upon concatenation, then it may be beneficial to not error correct each wait location, or to only error correct a wait location at higher levels of concatenation.  This scheme will still demonstrate a threshold; however the failure probability map will have to take into account the differences between the replacement rules for each type of location at different levels of concatenation.
By optimizing the frequency of error correction at different levels of concatenation and for different location types, the fault-tolerance threshold can be improved.

This work can be extended in several directions. First, we have put forward
a conjecture that pseudothresholds may lead to upper bounds on $\gamma_{th}$.
If this conjecture is true, then it would be interesting to determine how the concatenated flow map formalism could be modified to give rigorous bounds.
It would also be useful to determine how much a pseudothreshold can differ from the fault-tolerance threshold.

Second, the analyses presented here only account for Clifford-group gates. We did not analyze a Toffoli, $\pi/8$, or other nontrivial
gate needed for computational universality. It is still possible to apply the methods in this paper to those gates, but the corresponding
flow map component is more difficult to estimate well.
To determine a quantum fault-tolerance threshold, it is necessary to evaluate a computationally universal basis.  
With a universal basis, how much does the fault-tolerance threshold and sequence of pseudothresholds change?

\section{Acknowledgements}
\noindent

We would like to thank one of our referees for suggesting the relationship
between pseudothresholds and the frequency of error-correction operations.
Krysta Svore acknowledges support from an NPSC fellowship, and
Andrew Cross was supported by an NDSEG fellowship.

\nonumsection{References}

\end{document}